\documentclass[lettersize,journal]{IEEEtran}

\usepackage{amsmath,amsfonts}
\usepackage{array}
\usepackage[caption=false,font=normalsize,labelfont=sf,textfont=sf]{subfig}
\usepackage{textcomp}
\usepackage{stfloats}
\usepackage{url}
\usepackage{verbatim}
\usepackage{graphicx}
\usepackage[nocompress]{cite}
\usepackage{booktabs,tabularx}
\usepackage[hidelinks]{hyperref}
\newcolumntype{P}[1]{>{\raggedright\arraybackslash}p{#1}}
\newcolumntype{Y}{>{\raggedright\arraybackslash}X}
\newcommand{\figone}[3]{\begin{figure}[!t]\centering\includegraphics[width=\columnwidth]{figures/#1.pdf}\caption{#2}\label{#3}\end{figure}}
\newcommand{\figwide}[3]{\begin{figure*}[!t]\centering\includegraphics[width=\textwidth]{figures/#1.pdf}\caption{#2}\label{#3}\end{figure*}}

\hypersetup{pdfauthor={Heyue Li, Shaolin Xiang, Tian Tian, Qichun Liu, Tiefu Li, Zhihua Wang, Yanshu Guo, Hanjun Jiang},pdftitle={QCORE: A Quantum-Computing-Oriented, Resource-Efficient Runtime Architecture with Lifecycle Coordination}}

\begin{document}
\raggedbottom
\title{QCORE: A Quantum-Computing-Oriented, Resource-Efficient Runtime Architecture with Lifecycle Coordination}
\author{Heyue~Li, Shaolin~Xiang, Tian~Tian, Qichun~Liu, Tiefu~Li, Zhihua~Wang, Yanshu~Guo, and Hanjun~Jiang%
\thanks{This work was supported in part by the National Natural Science Foundation of China under Grants 62674147 and 92365210, the National Key Research and Development Program of China under Grant 2022YFA1405200, and the Beijing Engineering Research Center under Grant BG0149. \textit{Heyue Li and Shaolin Xiang contributed equally.} \textit{(Corresponding authors: Yanshu Guo and Hanjun Jiang.)}}%
\thanks{H. Li, S. Xiang, T. Tian, T. Li, Z. Wang, and H. Jiang are with the School of Integrated Circuits, Tsinghua University, Beijing 100084, China (e-mail: \href{mailto:jianghanjun@tsinghua.edu.cn}{\mbox{jianghanjun@tsinghua.edu.cn}}).}%
\thanks{T. Li is also with BNRist and Frontier Science Center for Quantum Information, Tsinghua University.}%
\thanks{Q. Liu is with the Beijing Academy of Quantum Information Sciences, Beijing, China.}%
\thanks{Y. Guo is with the School of Integrated Circuits, Shanghai Jiao Tong University, Shanghai, China (e-mail: \href{mailto:guoyanshu@sjtu.edu.cn}{\mbox{guoyanshu@sjtu.edu.cn}}).}}
\maketitle
\begin{abstract}
As quantum processors scale, runtime functions such as online calibration, quantum error correction, and measurement-driven feedback are becoming essential components of quantum control systems. These functions require controllers to continuously process classical data and update parameters for subsequent operations during program execution. To limit on-chip resource requirements and support scaling, we organize hardware around shared resources. In systems with shared classical compute resources, computation completion, the last parameter read, and result consumption generally occur at different times, requiring separate management of their dependencies while meeting the real-time requirements of quantum operations. This paper presents QCORE, a quantum-computing-oriented, resource-efficient runtime architecture that uses lifecycle coordination to enable safe reuse of shared compute resources and maintain parameter consistency at runtime. QCORE models resource occupancy, parameter protection, and consumer dependencies as independent lifecycles and introduces three coordination mechanisms: pre-use checks and parameter binding to verify timing, identity, and dependency conditions at execution; access-aware parameter protection using value snapshots, version references, or active-configuration windows according to object semantics; and joint admission of resource reservations and parameter installation to coordinate compute reuse with parameter-switching constraints. After request submission, hardware autonomously handles admission, protection, installation, and reclamation, reducing the need for fine-grained Host coordination of local execution and providing a foundation for one Host to manage multiple controllers. We implement synthesizable RTL and validate timing on an FPGA platform at 100 MHz. Independent release reduces whole-program execution time by 16.83\% in Ramsey calibration and QEC replay, while joint admission increases on-time completions from 5/15 to 15/15 under contention across five arrival phases. Under the same external deadlines, a shared NN backend reduces LUT usage by 12.28\%--13.81\% and DSP usage by 39.39\% relative to two statically assigned backends. These results show that hardware lifecycle coordination converts inter-task timing slack into opportunities for on-chip sharing, increasing service capacity under limited resources and supporting quantum control scaling.
\end{abstract}
\begin{IEEEkeywords}
Quantum computing, quantum control, runtime architecture, lifecycle coordination, resource efficiency, resource sharing, runtime parameter updates, quantum error correction, quantum calibration, deterministic execution, parameter consistency.
\end{IEEEkeywords}

\section{Introduction}
\IEEEPARstart{Q}{uantum} computing is progressing from small and medium-scale experimental platforms toward larger systems and longer, more complex program execution. As mid-circuit measurement, online calibration, and quantum error correction (QEC) become part of continuously running quantum programs, controllers must also perform classical processing, including measurement classification \cite{guoVLSI26,liASIC26}, parameter estimation \cite{liCal26}, and error decoding \cite{ref1,ref2,ref3}. These tasks must produce results in time while remaining compatible with ongoing pulses and measurements. For example, CaliQEC combines online calibration with error correction, further increasing runtime data-processing and parameter-update requirements \cite{ref2}. Frequent Host round trips for measurement processing and fine-grained coordination introduce communication, software scheduling, and data-movement overhead into the feedback path. As more tasks run concurrently, Host processing capacity may also limit the scale of control that it can manage.

Quantum operations impose two forms of timing requirements. Finite coherence times and gate requirements on pulse phase and spacing require pulse issuance, measurement capture, and multichannel synchronization to occur at specified times \cite{ref4,ref5}. Readout discrimination, feedback computation, and error decoding generally need to finish before subsequent operations consume their results. The system must therefore preserve deterministic timing for external events and, given limited on-chip resources and future scaling needs, exploit the slack of classical tasks to share computation and storage. Delegating runtime coordination to local hardware reduces the need for software intervention in these time-sensitive processes, while on-chip compute reuse improves hardware utilization.

Fig.~\ref{fig1} places QCORE between the Host and the QPU. Its left side shows the functional stack from applications, compilation, and logical operations through runtime processing and control/readout to the QPU. The Host generates and compiles programs, deploys tasks, and manages global experiments. QCORE combines local timed execution, shared calibration or decoding services, parameter management, and runtime coordination. The QCORE Quantum I/O Subsystem converts digital events into physical control/readout signals \cite{guoISSCC23,guoISSCC24}. Measurements may be processed locally or externally before subsequent use. One Host can manage multiple QCORE nodes, each coordinating a group of qubits after submission, while the Host handles program organization and cross-controller management.

This division complements NVQLink's low-latency cooperation between quantum systems and real-time CPUs/GPUs \cite{ref6}. NVQLink supports time-sensitive QEC and calibration computation; QCORE coordinates resource reuse and parameter activation within a controller. Tasks may execute locally or externally according to scale, data location, and deadlines. Returned results must still satisfy scheduled operations' parameter and timing requirements, linking compute cooperation to local execution.

Existing quantum control microarchitectures and FPGA platforms provide explicit timing, programmable feedback, and heterogeneous compute interfaces \cite{ref4,ref5,ref7,ref8,ref9,ref10}. Further sharing of classical processing resources introduces a cross-stage problem: parameter installation, subsequent reads, and result output may continue after a compute unit finishes. For example, a calibration estimator can process another dataset after producing candidate parameters, while scheduled old operations still need the original parameters and new parameters must take effect before subsequent use. Similarly, after a decoder exits, its results may still await a Pauli-frame update or conditional operation. Retaining resources until the entire service finishes extends hardware occupancy through consumer waiting periods. Conversely, accepting new tasks based only on local idleness can invalidate an existing parameter-switching plan. Updates competing for the same installation port also require a jointly feasible schedule.

QCORE organizes execution around these dependencies. Pre-use checks verify timing and context at actual execution. Parameter management protects old operations according to three access modes: value snapshot, version reference, and active configuration. Joint admission checks both resource reservations and legal installation slots for pending updates. Resource exit, the last parameter read, installation acknowledgment, and output completion each release their corresponding dependencies, enabling safe local hardware reuse before the service finishes. Hardware handles this coordination autonomously after request submission, moving per-operation lifecycle management into the controller.

This paper makes three contributions. First, it establishes a hardware-executable model separating resource use, parameter installation, and subsequent consumption while protecting registered future consumers, providing explicit objects and rules for control-plane offloading. Second, it introduces a hardware protocol connecting admission, parameter management, and actual endpoints. Joint feasibility checks, event feedback, stale-proposal rechecking, and independent reclamation preserve existing commitments when new work arrives. Third, it implements synthesizable RTL and quantifies safety, on-time service capacity, and hardware cost through real-endpoint regression, same-resource policy ablations, parameter-size sweeps, conventional-mechanism comparisons, and digital Service synthesis.

Independent release reduces program completion time by 16.83\% in Ramsey calibration and QEC replay. Joint installation planning increases on-time completions from 5/15 to 15/15 in the five-phase experiment. A shared NN backend reduces LUT usage by 12.28\%--13.81\% and DSP usage by 39.39\% under the same external consumption schedule. Section~II reviews related work; Sections~III and IV present the architecture and mechanisms; Section~V evaluates the implementation; Section~VI discusses scale and related systems; and Section~VII concludes. The separate appendices provide baseline distinctions, feasibility analysis, and interface details.

\figwide{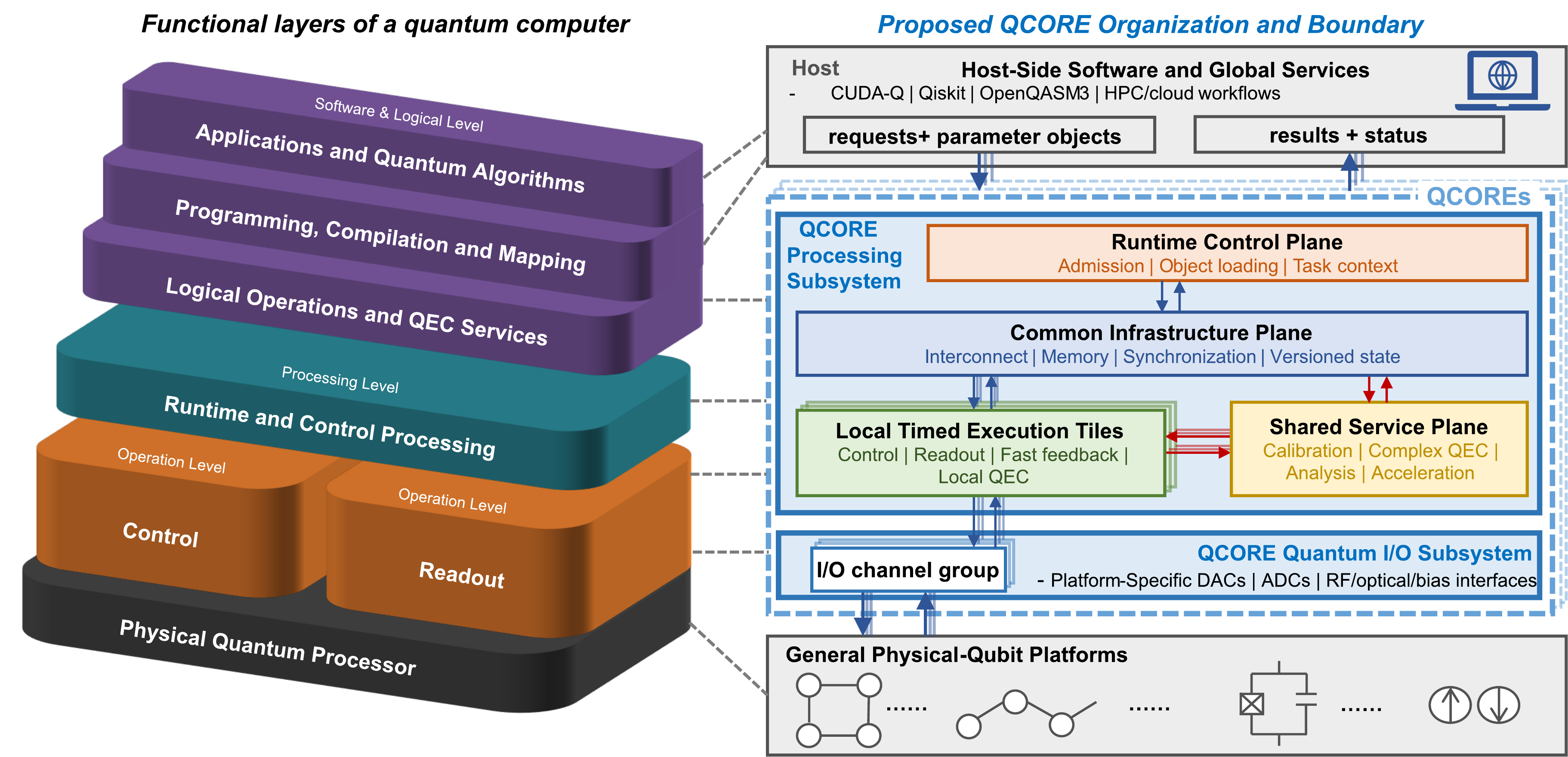}{Functional organization of a quantum computer and QCORE. The left shows the general functional stack. On the right, moving local computation and lifecycle coordination onto QCORE reduces fine-grained Host involvement, enabling one Host to manage multiple QCORE nodes. QCORE integrates timed control, readout processing, error correction, and calibration within an on-chip architecture, supporting greater integration and scalability of quantum computing systems. The Host retains program organization and global management, while the QCORE Quantum I/O Subsystem connects the QCORE Processing Subsystem to the QPU.}{fig1}

\section{Background and Related Work}
\subsection{Timing Constraints and Future Consumers in Quantum Execution}
Quantum programs combine deterministic physical events with deadline-constrained classical processing. Pulse issuance, measurement capture, and synchronization must satisfy predefined timing requirements, whereas the classical results derived from measurements are constrained by when subsequent operations use them \cite{skoric23}. OpenQASM~3 and surface-code execution models \cite{ref1,ref11,ref12} reflect both timing constraints and classical dependencies in quantum programs.

QEC illustrates this temporal separation. Repeated measurements produce error information that a classical decoder processes \cite{afs22}, and the decoding results are consumed through Pauli-frame updates or feedback operations. Decoding may finish before downstream consumers complete, creating opportunities to release and reuse compute resources early.

Online calibration further introduces runtime state changes \cite{kelly16}. Measurements and estimation produce candidate parameters, which must be loaded and installed before subsequent operations can use them. Meanwhile, previously scheduled operations may still access the old configuration, giving old-parameter retention and new-parameter activation distinct timing requirements.

\subsection{Quantum Control Architectures with Deterministic Timing}
Existing quantum control architectures provide deterministic event execution, programmable feedback \cite{ryan17}, and scalable interfaces. Instruction architectures such as QuMA and eQASM support quantum operations through explicit timing control and execution scheduling \cite{ref4,ref5,ref7}.

FPGA platforms such as QICK, QubiC, and RISC-Q integrate pulse generation, readout processing, and programmable control \cite{ref8,ref9,ref10,ref22,ref26}; distributed synchronization and timestamp-based execution extend these capabilities to larger systems \cite{ref13,ref14,ref15}. These architectures support deterministic execution of submitted programs. Cooperation with shared calibration or decoding services additionally requires runtime coordination of reservations, parameter visibility, and consumers that outlive computation, which QCORE implements in controller hardware.

\subsection{Shared Classical Computation and Runtime Parameter Updates}
Classical processing loads in quantum control grow with qubit count, program complexity, and measurement frequency. Recent work on elastic QEC decoding \cite{ref16} shows that limited decoder resources can be allocated dynamically according to task demand. Such approaches exploit the slack between computation completion and result use to share compute capacity while satisfying consumer deadlines.

Runtime calibration further requires simultaneous management of multiple parameter states. Hardware endpoints may access parameters by directly reading values, accessing versioned storage, or reading active configuration registers. These access modes require different protection mechanisms so that old operations continue to observe consistent data while new operations obtain updated configurations.

Runtime revision mechanisms, such as those in time-deterministic and instruction-dynamic execution (TIDE) \cite{ref17}, provide protection boundaries for adaptive quantum execution. Shared-backend execution adds another requirement: parameter and consumer dependencies may remain after compute resources become idle. QCORE therefore manages compute-resource release, parameter-state retention and updates, and subsequent operation completion separately, enabling safe local reuse before the entire service ends.

\subsection{Relation to Conventional Dependency Management and Real-Time Scheduling}
Conventional computing systems have mature mechanisms for dependency management and real-time scheduling. Reorder buffers (ROBs) support in-order instruction retirement and precise state recovery \cite{ref18}. Read-copy update (RCU) separates publication of a new version from reclamation of old objects, allowing existing readers to continue accessing the original objects after an update \cite{ref19,ref28}. Real-time scheduling analyzes schedulability using task deadlines and processing capacity \cite{ref20}. These methods provide foundations for identity management, version protection, and timing analysis in QCORE.

The quantum control problem is to make these mechanisms jointly satisfy endpoint execution contracts. Physical operations have scheduled occurrence times; registered but not yet issued operations may still need old parameters; and candidate parameters must be installed before subsequent use. When an old object can be reclaimed and when a new configuration must take effect are distinct constraints. Multiple updates sharing an installation port also need a jointly feasible execution schedule.

QCORE records resource occupancy, parameter references, and consumer completion separately. It incorporates future reads and installation deadlines into the same admission check and drives reclamation with actual endpoint events. Its contribution lies in the execution model, cross-module protocol, and synthesizable hardware implementation for these constraints. Appendix~A uses comparisons with identical capacity and timing constraints to distinguish the effects of reclamation granularity and installation-plan revision, and independently checks joint-planning feasibility against an exhaustive reference.

\section{QCORE Architecture and Lifecycle-Aware Execution Model}
\subsection{System Boundary and Interfaces}
The QCORE Processing Subsystem is a QPU-side digital processing system connecting deterministic quantum control execution with shared classical compute resources. It comprises the Runtime Control Plane, Common Infrastructure Plane, Local Timed Execution Tiles, and Shared Service Plane. Together with the QCORE Quantum I/O Subsystem, it forms a QCORE node. The Host supplies quantum programs and configuration data to the Processing Subsystem; the Quantum I/O Subsystem connects it to platform-specific control and readout components. The implementation and evaluation in this paper primarily cover coordination logic, parameter endpoints, and shared compute backends within the digital Service. System-level pulse generation, the readout frontend, and Host program management connect through the defined timing and data interfaces.

The Host or an upstream control unit submits requests over deployment-specific links, such as PCIe, Ethernet, or other interfaces. A request specifies the target operation, timing requirements, required inputs, and result destination. After checking resources and dependencies, the coordinator forwards it to an execution endpoint and maintains subsequent state from hardware events. Deterministic control and capture interfaces transmit timestamped events through the Quantum Timing Interface (QTI). Transport latency, buffering, and bounded backpressure are included in the timing budget. For example, with interface latency $\ell$, an event required at time $T$ must be delivered by the digital subsystem no later than $T-\ell$. One Processing Subsystem connects to one logical Quantum I/O Subsystem, which may contain multiple I/O modules. The numbers of execution tiles, I/O modules, and DAC/ADC channels are configured separately. Supported qubit count also depends on modulation, readout multiplexing, and concurrent-operation requirements; these quantities do not have fixed one-to-one correspondences. Fig.~\ref{fig2} illustrates grouped scaling with $S$ QCORE nodes, $K$ groups of execution tiles and I/O channels per node, and $n$ qubits served by each group. The product $S\times K\times n$ represents the organizational scale under a given channel mapping; DAC/ADC counts still depend on each group's control and multiplexing scheme.

\subsection{Hardware Organization}
Fig.~\ref{fig2} shows the system organization and the placement of the proposed coordination mechanisms. The Runtime Control Plane maintains task contexts and admission state, local timing logic issues events as scheduled, and shared services perform classical computation. System components in the figure indicate interface responsibilities; synthesis and endpoint validation are bounded by the digital Service top level. Use-Point Protection is distributed across issue, parameter-read, and result interfaces, whose implementations and signal connections are detailed in Figs.~\ref{fig4} and \ref{fig5}.

The Runtime Control Plane and Common Infrastructure Plane manage program state, request scheduling, data transfer, and execution status. They include the Host interface, program and parameter staging, Management CPU, Task context table, Admission FSM, on-chip memory and interconnect, a common time base, and parameter-version management. The Management CPU organizes program phases, initiates services, and handles status; local timing hardware generates cycle-level events. Task contexts hold operation identities, required objects, and result destinations. Hardware coordinates resource reservations, parameter protection, installation, and independent reclamation after request submission.

Runtime information is divided into bulk data and short control events. Programs, parameter objects, and measurement data use memory or data interfaces. Reservations, actual exits, last reads, state installation, and completion responses update management state. In Fig.~\ref{fig2}, the Object Interconnect carries program and parameter objects; the Control Interconnect carries requests, grants, and lifecycle events; and the Measurement Interconnect carries measurement and service data. The implemented Service connects adapters, the ledger, parameter management, and execution endpoints through direct ports and local queues.

\figwide{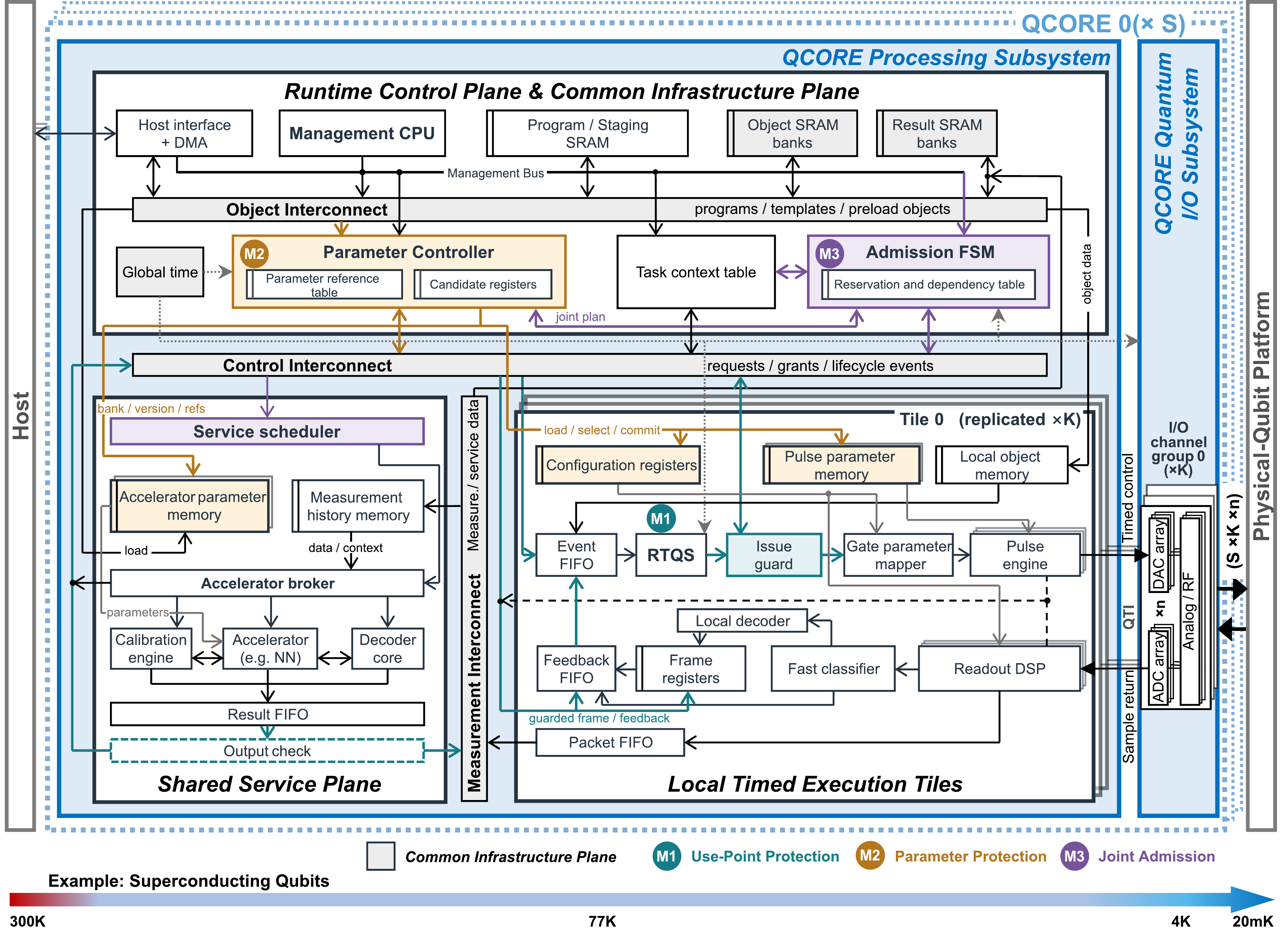}{QCORE hardware organization and logical interfaces. Runtime Control Plane manages programs, tasks, resource admission, and parameter updates; Common Infrastructure Plane provides distributed storage, interconnects, and a common time base (gray hardware); Local Timed Execution Tiles perform timed control, readout processing, and fast feedback; Shared Service Plane provides reusable computation for calibration, decoding, and other classical tasks. M1--M3 denote Use-Point Protection, Parameter Protection, and Joint Admission. $S$, $K$, and $n$ denote node count, execution/I/O channel groups per node, and qubits per group. Section~III-A defines the digital Service boundary; the temperature band illustrates functional distribution in a superconducting platform.}{fig2}

Local Timed Execution Tiles handle deterministic control, readout processing, and fast feedback. They contain event queues, time comparison and issue logic, a Gate parameter mapper, digital pulse generation, and readout-processing modules. Operations with strict low-latency requirements can execute locally, whereas tasks needing more computation or historical data are sent to the Shared Service Plane.

The Shared Service Plane supplies execution resources for calibration estimation, decoding, statistical processing, and other classical tasks. It includes request scheduling, measurement and history-data management, compute-resource allocation, and result-return interfaces. The scheduler allocates compute opportunities according to resource state and task deadlines. Results and update state connect to subsequent operations through interfaces with explicit identity, timing, and parameter contracts. The neural-network (NN) pipeline is one shared compute endpoint used in this paper. Lifecycle contracts are independent of the algorithm, allowing other local or external accelerators to declare occupancy and dependencies in the same way.

\subsection{Resource, State, and Consumer Model}
A service in QCORE is a group of operations organized around a goal, such as estimating parameters and applying the updated parameters in subsequent quantum operations. A service may have multiple consumers and may share compute resources or parameter objects with other services. Parameter objects include classical data such as frequency, phase, waveforms, classification coefficients, and decoding contexts. Their access modes determine their protection methods.

\figone{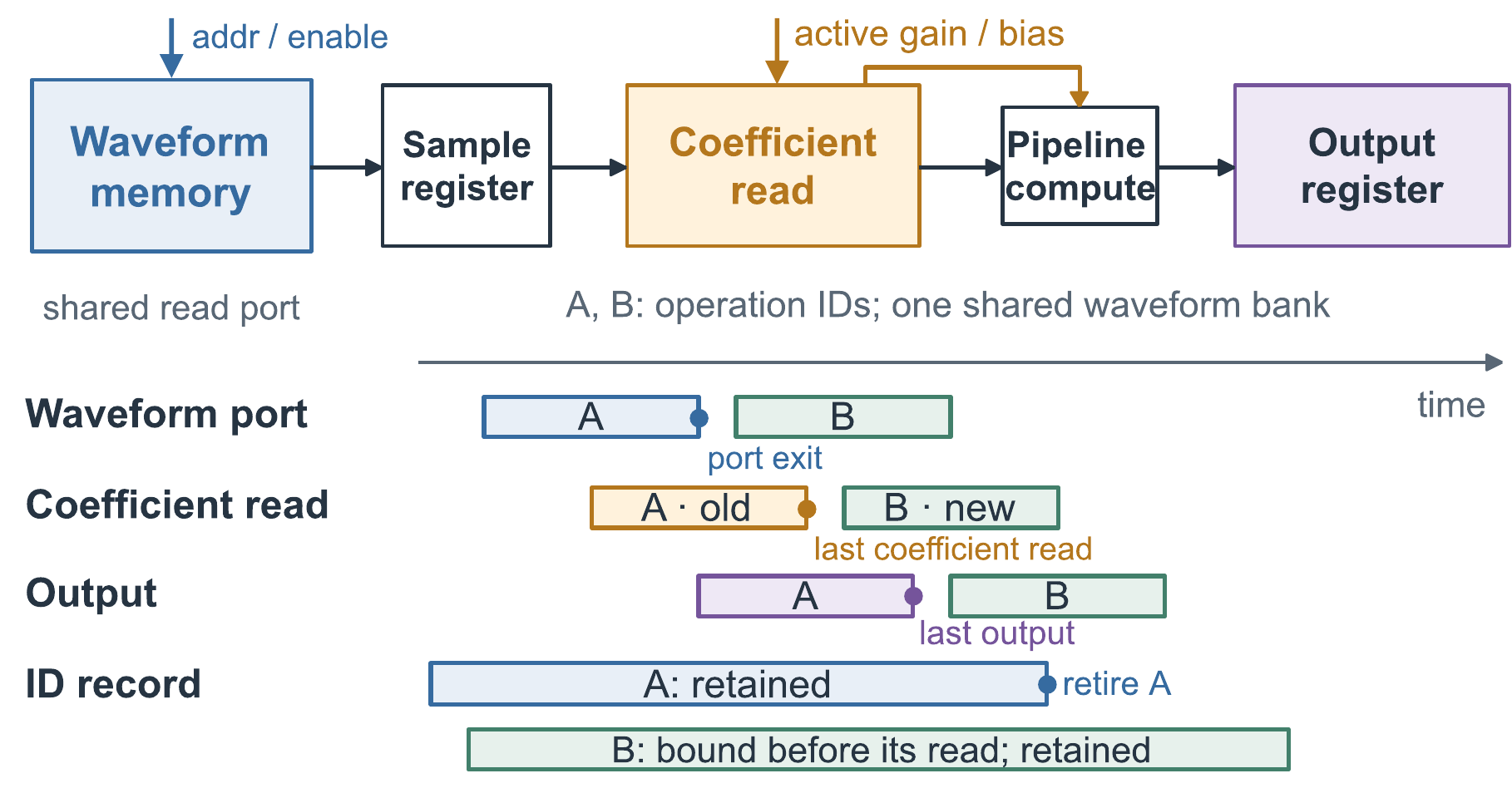}{Lifecycle separation within one waveform and coefficient-processing endpoint. The top shows the shared waveform read port, sample register, coefficient read, compute pipeline, and output register. The bottom shows port use, coefficient reads, output, and identity-retention intervals for A and B. Port exit, the last coefficient read, the last output, and identity retirement release different dependencies. Operation B can use the shared port before A finishes downstream processing. A and B reference the same waveform bank but use the old and new active coefficients, respectively. The timeline indicates event ordering.}{fig3}

QCORE maintains five independently released records. Resource reservations contain the resource, interval, and operation ID, and end on exit or cancellation. Update records hold the candidate, version, window, and status until installation or failure. Parameter-reference records retain bound values or versions and reference counts through the last read. Consumer records track sources, access windows, deadlines, and completion until consumption or output finishes and responses drain. Identity records retain the ID, generation, and pending responses until all dependencies complete. A resource can therefore accept new work after exit while records protecting subsequent reads and outputs remain valid.

Fig.~\ref{fig3} illustrates these boundaries within one endpoint. A shared waveform-bank read port reads samples cycle by cycle. Registered samples enter a multiply--accumulate pipeline that reads the active gain and bias and then produces outputs. After the read port finishes operation A, it can serve B while A's downstream stages may still read coefficients and compute results. Thus, the last waveform read, last coefficient read, and last result output are distinct events, and the identity record remains until associated responses drain. A and B denote operation identities and may reference the same waveform bank; their old and new active coefficients belong to a separate parameter object. Waveform and coefficient references are released at their respective last reads. If another operation still references the same waveform version, that version remains retained. References are established at binding and cover accepted but not yet issued operations, preventing required data from being overwritten during queueing. A similar separation occurs at service level: the calibration estimator or decoder can exit after computation, while the corresponding records continue to track parameter installation and subsequent consumption.

\subsection{Timing Model and Admission Conditions}
QCORE uses a common discrete-time model to coordinate resource reservations, parameter installation, and consumer deadlines. For a resource of capacity one, operation $i$ reserves resource $r_i$ over the half-open interval $[s_i,e_i)$. Distinct operations $i$ and $j$ must satisfy
\begin{equation}\label{eq1}
r_i=r_j\ \Rightarrow\ (e_i\le s_j\ \lor\ e_j\le s_i).
\end{equation}
Pipeline input ports, memory interfaces, and internal compute resources are modeled separately according to their actual release boundaries.

Consumer $r$ declares a set of legal start times $I_r$, a conservative start-to-completion budget $B_r$, and a result deadline $d_r$. For an exactly timed event, $I_r$ contains a single time; schedulable operations may have multiple legal start times. Acceptance or use of an operation at time $t$ requires
\begin{equation}\label{eq2}
t\in I_r,\qquad t+B_r\le d_r.
\end{equation}
Inputs must also be ready, resource capacity sufficient, and parameters valid.

A request proceeds through enqueueing, ledger authorization, and endpoint acceptance. It may wait or be canceled after enqueueing. Ledger authorization registers the resource reservation, operation identity, and installation plan. The adapter and endpoint then establish parameter references through handshakes and check timing and parameter conditions before execution. If downstream binding or acceptance fails, identity-tagged exit and retirement events clean up registered records. The execution budget includes endpoint processing latency and bounded backpressure permitted by the interface contract. An operation ID and generation jointly identify an operation instance, preventing late responses from an old operation from modifying a reused slot. Parameter versions and operation generations are managed separately. The current parameter banks block further publication once the 8-bit version reaches its limit; related references and responses must drain before this version space is reused.

\section{Runtime Coordination Mechanisms}
QCORE preserves deterministic quantum control timing and parameter consistency by managing resource occupancy, parameter protection, and consumer dependencies separately. Pre-use checks and binding, access-aware parameter protection, and joint admission of reservations and installation retain the state needed after resource exit and prevent new work from invalidating registered plans.

\subsection{Pre-Use Checks and Parameter Binding}
Quantum execution plans are generally generated in advance. Before execution, queued requests may encounter time advancement, cancellation, or configuration changes. Endpoints therefore recheck conditions at actual issue. The Issue guard checks task and round, required configuration version, time validity, execution budget, and resource authorization. The adapter and ledger match the complete operation identity. This connects the feasibility decision at registration with condition checking at actual execution.

A valid request issues when the downstream endpoint is ready. A valid but temporarily blocked request waits. A request with invalid identity, version, or timing conditions is rejected and enters the corresponding cleanup path. Adapter--endpoint handshakes establish parameter references, which remain in the operation context until the last read of the corresponding object. Accepted but not yet issued operations retain references as well, protecting queueing, request transfer, and actual access.

The active-coefficient endpoint compares expected and actual versions at the read edge. It propagates data, identity, version, first/last flags, and a per-sample error bit together to the Service boundary, and records mismatches in a sticky \texttt{fault}. Application consumers may use only valid, error-free samples; atomic acceptance of an entire operation additionally requires accumulated error state and buffering or deferred commitment. Last-read and last-output events still release references and drain identities for erroneous operations. Thus, detecting an error, accepting a result, and cleaning up the operation remain separate responsibilities (Appendix~C).

\subsection{Parameter Protection and Installation Windows}
Runtime parameters may be captured before execution or read continuously from banks or active registers. QCORE matches protection to these access modes and their actual use times.

For small parameters determined at acceptance, a value snapshot copies the parameter into operation-private state. Once capture completes, the source may change while the operation continues using its private value until completion.

For waveforms, weights, and other large objects, version binding avoids per-operation copying costs. Candidates load into separate storage and are published only after preparation completes. New operations bind the published bank and version, while existing operations retain their old selection through the final read. Storage can be reused only when inactive and unreferenced.

For endpoints that directly read a set of active configuration registers during execution, old and new operations access the same active state, requiring coordination between reads and configuration switching. QCORE defines an installation window so that the new configuration takes effect after protected old-state reads, while meeting the update deadline and the first-use requirements of registered new consumers. For update $u$, let $S_u$ be the discrete commit slots permitted by the endpoint and $p_u$ the first eligible commit edge after the candidate is fully prepared. Let $O_u$ be the set of registered old-state consumers, and $\bar r_j$ the upper bound on the last read of operation $j$ in that set. Bound operations that are still waiting are included in $O_u$. If the endpoint must freeze new acceptance of old-state operations $q_u$ cycles before installation, let $\bar a_j$ be the latest acceptance time already committed to old consumer $j$. Then $T_u-q_u>\bar a_j$ is also required so that an early freeze does not remove its acceptance opportunity.

Let $N_u$ denote registered new-state consumers, $c_k$ the first-use time of operation $k$ in that set, $\delta_u$ the delay from installation to new-state visibility, and $x_u$ the update's latest commit time. Set $\sigma_u=0$ if reads on the installation edge observe the pre-edge old value, or $\sigma_u=1$ if installation must occur strictly after the last old read. The installation window is defined by
\begin{equation}\label{eq3}
\begin{aligned}
L_u&=\max\!\left(p_u,\max_{j\in O_u}(\bar a_j+q_u+1),\right.\\[-2pt]
&\hspace{12mm}\left.\max_{j\in O_u}(\bar r_j+\sigma_u)\right),\\
U_u&=\min\!\left(x_u,\min_{k\in N_u}(c_k-\delta_u)\right),\\
W_u&=S_u\cap[L_u,U_u],\qquad T_u\in W_u.
\end{aligned}
\end{equation}
Terms taking a maximum or minimum over an empty set are omitted. The acceptance-bound term can be omitted for endpoints that do not require an early freeze. Fig.~\ref{fig5}(b) combines candidate readiness, old-operation acceptance, and last reads into the lower bound, and first new-state use and the update deadline into the upper bound. In the figure, $c_u$ denotes the earliest new-state use. $S_u$ captures additional endpoint-specific commit restrictions. The active-coefficient endpoint in this paper requires no early freeze: reads on the installation edge use the old value, and reads on the next edge use the new value. Thus, $\sigma_u=0$ and $\delta_u=1$, and preparation completion and the last old read determine its lower bound. At installation, the adapter rechecks candidate completeness, plan certification, arrival of the scheduled time, and old-reference safety before switching the active coefficients and reporting completion. New dependencies can change legal windows. If a window becomes empty or no joint schedule exists with other updates, the new request cannot join the current plan.

\figone{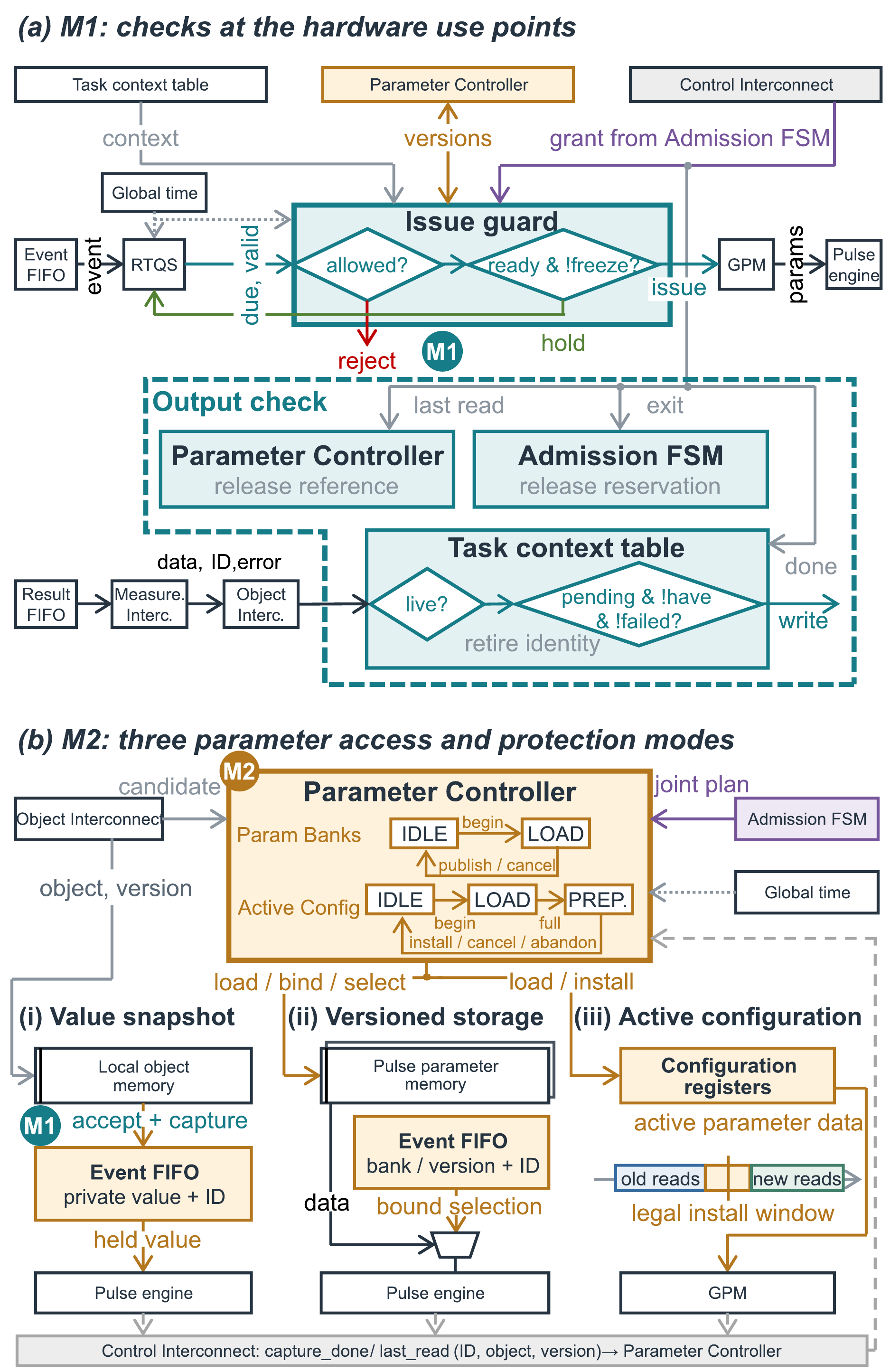}{Hardware paths for use-point and parameter protection. (a) M1: RTQS forwards due events to the combinational Issue guard, which checks task context, parameter bindings, admission authorization, and current time. It issues valid requests when downstream is ready, holds temporarily blocked requests, and rejects invalid requests. Results carry identity and error information. Parameter Controller releases object references after last read; Admission FSM releases reservations at actual resource exit; Task context table reclaims identities after dependencies and responses drain. (b) M2: Parameter Controller loads candidates through Object Interconnect. Snapshots retain operation-private copies after capture; version binding records bank/version selections to protect queued and executing reads; active configuration waits for a legal installation time.}{fig4}

One operation can combine multiple parameter-protection modes. For example, the amplitude in a descriptor can use a snapshot; waveform data can use a bank and version reference; and downstream gain and bias can be updated through an active-configuration window. Actual endpoint access semantics determine the method: snapshots require private copies, versioned access requires object selection and retention, and active configuration requires checking the switching time. Fig.~\ref{fig4} connects these methods to issue, read, and output interfaces. Fig.~\ref{fig5} shows how installation timing and admission plans are maintained jointly.

\subsection{Joint Admission of Resource Reservations and Parameter Installation}
Checking resource idleness and parameter validity separately is insufficient to establish whether a new request can coexist with an existing execution plan. Even if a task obtains a compute interval, it may extend old-parameter reads, advance new-parameter use, or compete with other updates for the installation port. QCORE therefore checks resource reservations and installation feasibility for affected updates in the same admission decision, as shown in Fig.~\ref{fig5}(a).

When a request arrives, the coordinator first checks available execution intervals against existing reservations. It then recomputes installation windows affected by the new consumer and checks whether they jointly satisfy the shared commit port's capacity. The request adapter, admission decision core, and Commitment Ledger perform this process together. The decision core forms a proposal from reservations, windows, and a state revision number. After the ledger performs final checks and registers resources and the plan, the adapter establishes references and delivers the request to the endpoint. Execution requires available resources, valid parameter conditions, and jointly feasible updates.

Exit, retirement, and installation events immediately update the relevant records and increment the state revision number. A proposal based on stale state becomes invalid and is rechecked as required. Final registration checks the revision number and timing conditions again. The ledger atomically registers reservations, identities, and the plan; endpoint references are then established through handshakes. Identity-tagged exit and retirement events clean up authorized requests that fail downstream acceptance.

For a shared commit port, the coordination core follows a fixed update order and selects each update's earliest legal slot at least one cycle after the preceding commit (Fig.~\ref{fig5}(c)). With one port, one-cycle commits, and fixed legal sets, this constructs a schedule whenever one exists in that order. Once legal sets are formed, sequential selection examines at most $\sum_i k_i$ candidate slots, whereas exhaustive enumeration visits $\prod_i k_i$ complete vectors, where $k_i$ is update $i$'s legal-slot count. These different work units describe algorithmic growth. Appendix~B gives the proof and conditions.

A legal window records the slots available to one update. The joint check additionally considers all relevant updates, shared-port capacity, and resource reservations to determine overall feasibility. A new old-state consumer may move an installation later, provided every registered constraint remains satisfied. If no joint schedule exists, the coordinator rejects the new request and retains existing commitments, allowing additional reuse only when execution conditions remain feasible.

\figwide{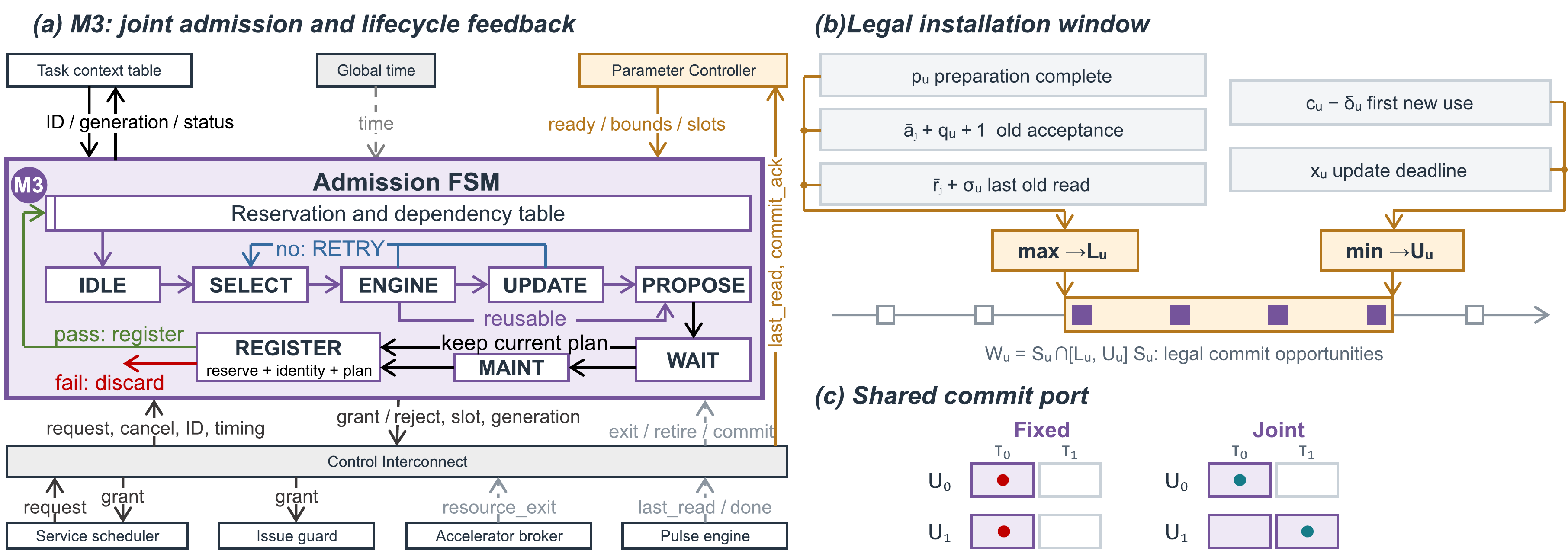}{Joint admission, installation-window formation, and shared commit-port coordination. (a) Admission FSM checks resource intervals and forms joint installation plans from task identity, current time, reservations, and parameter dependencies. After rechecking identity, time, and state revision, it atomically registers reservations, identity, and plan before returning authorization downstream. Actual exit, retirement, and installation events update the ledger. Stale proposals are rechecked; infeasible requests are rejected without changing existing commitments. (b) Candidate readiness, committed old-operation acceptance times, and last old-parameter reads set the lower bound; first new-parameter use and the update deadline set the upper bound. Intersecting the bounded interval with endpoint-permitted commit slots defines the legal window. (c) Individually legal windows can still yield competing fixed candidate times on one port. Joint planning selects distinct one-cycle commit slots in a fixed update order, satisfying both updates' constraints.}{fig5}

\subsection{Independent Release, Safety Guarantees, and Endpoint Integration}
During execution, each completion event releases its corresponding dependency. Actual compute-unit or port exit releases the resource reservation. The last read of a parameter object releases its reference. Installation acknowledgment ends the update transaction. Consumer or output-stream completion advances result-state and identity-record cleanup. The operation identity is reclaimed only after all associated dependencies and responses have drained. Upstream resources can thus be reused before the service finishes, while downstream stages retain the parameters and output contexts they need.

To prevent state errors from lifecycle separation, QCORE checks the operation ID and generation of completion, cancellation, and read events. Only an event matching the current operation instance can modify runtime state, preventing late responses from affecting reallocated resources.

Under the declared timing budgets and interface constraints, resource checks preserve reservation capacity, parameter protection preserves the data state of bound operations, joint admission maintains existing-plan feasibility under new dependencies, and identity matching isolates operation instances. Endpoints must correctly report actual exit, last-read, and completion events; result receivers must follow the error-flag and result-acceptance contract. Explicit events connect these interface responsibilities, making lifecycle separation realizable in actual endpoints.

QCORE connects different control and compute units through a common endpoint interface. An endpoint supplies its resource-occupancy interval, parameter-access mode, execution budget, and consumer information, and reports actual exit, last parameter read, and completion. These lifecycle constraints allow the coordinator to support calibration, decoding, and other shared compute tasks independently of their algorithms.


\section{Experimental Evaluation}
This section evaluates execution safety, on-time service under limited resources, parameter-protection tradeoffs, and shared-backend cost under common external deadlines.

\subsection{Platform and Methodology}
The digital implementation is synthesized for a Xilinx xc7z020clg400-1 FPGA, with timing validated at 100 MHz. The experimental integration contains a shared NN pipeline, a waveform-read endpoint, downstream active gain/bias registers, and QCORE's admission, parameter-management, and lifecycle-reclamation logic.

Validation is organized by execution boundary. The actual Service RTL covers waveform and active-coefficient endpoints. Ramsey calibration and QEC programs combine software arithmetic with RTL coordination replay. Multiple updates sharing a port are evaluated through independent coordination-core replay. Conventional reclamation and installation policies are compared using executable models under identical constraints. The main Service integrates one active-configuration object. An independent coordination core handles the two-update example, with endpoint timing constraints supplied by software and logical time held constant during checking, to validate admission and plan maintenance. Model decision steps and RTL control cycles are counted separately.

The same-resource ablation includes three safe policies. A0 holds resources until the entire service ends and keeps the initial installation time fixed. A1 releases resources independently upon actual exit but retains the fixed initial installation time. A2 combines independent release with joint window planning. All three share parameter protection, identity management, endpoint checks, and exception cleanup; only release boundaries and installation policies differ.

Comparisons keep compute capability, storage and record capacities, and external deadlines identical. Whole-program experiments use the same program, seeds, and computation, although earlier feedback may affect arrival times. Fixed-arrival experiments additionally use identical request timestamps. Completion fractions use all offered requests as the denominator; a request counts as complete only if it satisfies the relevant execution and update constraints.

\subsection{Correctness of Lifecycle Separation}
We first verify safe overlap between resource release, subsequent parameter accesses, and the result pipeline. We record port exit, parameter reads, output completion, and identity reclamation in the actual Service RTL, and check output values and parameter versions.

Operation A is accepted in cycle 18 and reads the waveform in cycles 59--62. Operation B is accepted in cycle 48 and waits before using the same read port in cycles 63--66. A's resource exit is acknowledged in cycle 63, but its last active-coefficient read occurs in cycle 64, its last output in cycle 66, and its identity retires in cycle 68. The configuration is installed in cycle 64: A's last read on that edge still observes the old coefficients, while B reads the new coefficients from cycle 65 and finishes output in cycle 70.

Both A and B bind to waveform bank 0. Their respective last reads decrement the reference count, and the bank remains protected until B also finishes reading. This trace shows that port reuse, object protection, and the result pipeline have independent boundaries that permit execution overlap while preserving parameter consistency.

Table~\ref{tab3} summarizes endpoint-regression and bounded-model-checking coverage. Actual endpoints are checked for values, versions, event timing, and identity isolation. The listed 44/60-step bounded checks target the corresponding coordination models. Exception cases cover cancellation, incomplete loading, and changes during admission, all reaching the expected terminal states. Appendix~C defines the implementation scope and interface contracts for these forms of evidence.

\begin{table}[!t]\caption{Safety Validation}\label{tab3}
\centering\footnotesize
\begin{tabularx}{\columnwidth}{@{}P{.23\columnwidth}YY@{}}\toprule
Property & Checks & Evidence\\\midrule
Issue validity & Timing, grant, source & Endpoint/admission RTL\\
Parameter consistency & Version, last read & NN, waveform, coefficient RTL\\
Identity isolation & Cancellation, late/duplicate events & Event/terminal-state checks\\
Bounded safety & Reads, concurrent transactions & Coordination models: Z3, 44/60 steps\\\bottomrule
\end{tabularx}
\par\smallskip\raggedright All checks passed.
\end{table}

\subsection{Execution Benefits of Independent Release and Joint Admission}
\subsubsection{Resource Overlap in Whole Programs}
To evaluate independent release in whole programs, Ramsey calibration and QEC replay compare whole-service hold with independent release under identical programs, seeds, computation, and deadlines. Parameter-estimation and decoding arithmetic run in software, while coordination is verified by replay through the actual RTL. Fig.~\ref{fig6}(a) shows computation overlapping preceding consumers, and Fig.~\ref{fig6}(b) summarizes additional holding, reuse, and program completion.

Both policies complete 14 jobs and 3,200 cycles of useful computation. Whole-service hold retains the resource for another 10,610 cycles after computation. Independent release eliminates this post-compute hold and enables three reuses of the shared engine while old consumers remain live, totaling 1,200 cycles of overlapping useful computation. Program completion time decreases from 20,163 to 16,769 cycles, a reduction of 16.83\%. Cumulative resource occupancy decreases from 15,388 to 4,766 cycles; this metric also includes changes in reservation holding before computation.

Fig.~\ref{fig6}(a) shows a representative interval. The old calibration computation runs in cycles 1569--1669, and its consumer remains live until cycle 3021. Independent release allows subsequent QEC computation in cycles 2388--2788, overlapping the old consumer for 400 cycles. Whole-service hold releases the resource in cycle 3022 and delays QEC computation to cycles 3153--3553. Full-recheck and ordinary incremental implementations with the same independent-release capability also finish in 16,769 cycles. Thus, this benefit comes from separating actual resource exit from the consumer's terminal state.

\figone{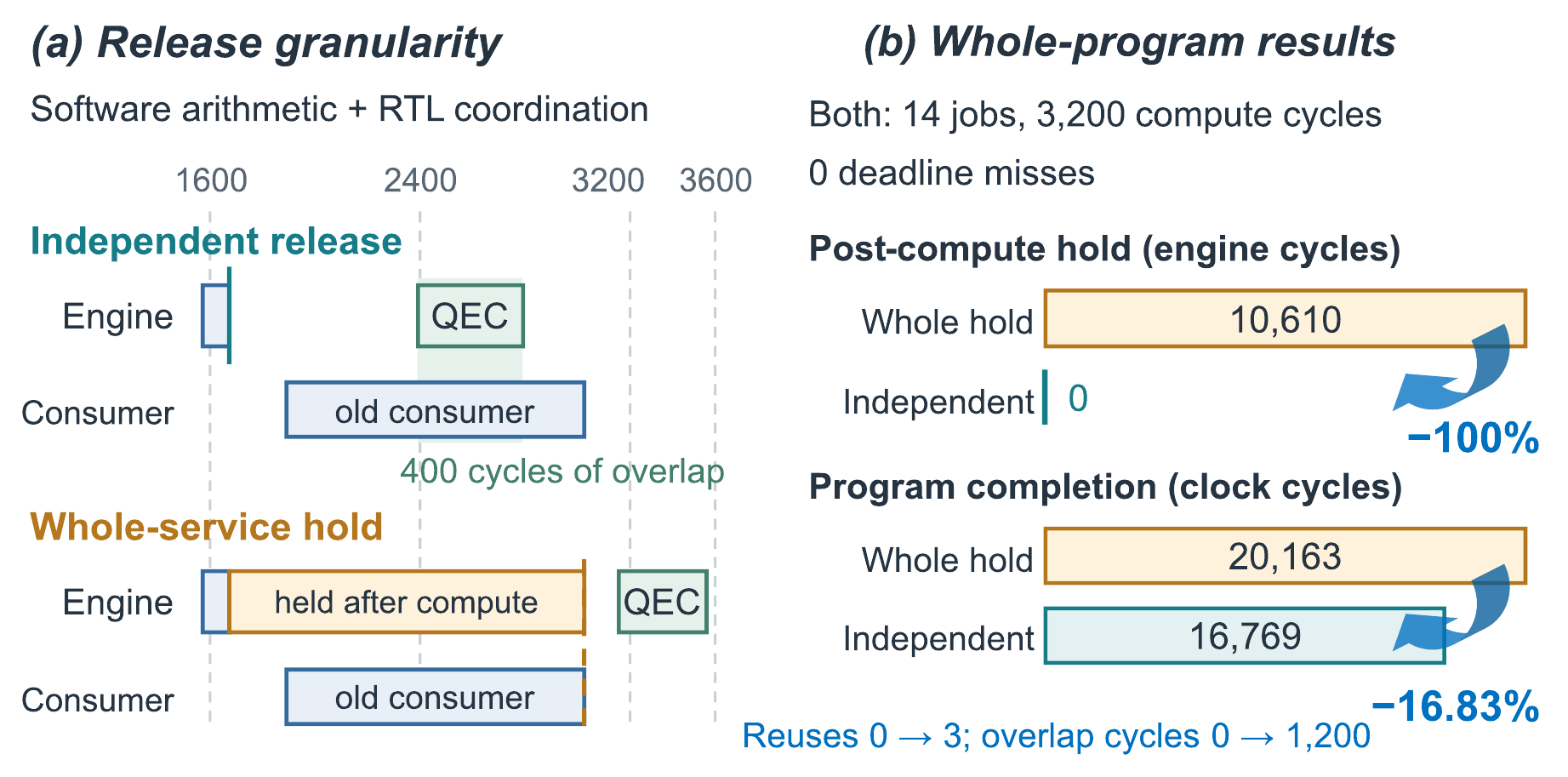}{Program-level benefits of independent resource release. (a) Shared-engine and old-consumer intervals under whole-service hold and independent release. (b) Additional holding, reuse, overlap, and program completion for the same 14 jobs and 3,200 cycles of useful computation. The experiment combines software arithmetic with actual RTL coordination replay.}{fig6}

\subsubsection{Release Benefits Under Fixed Arrivals}
To remove the influence of changes in dynamic program arrivals, both policies receive exactly the same request timestamps. At offered compute loads of 20\%, 40\%, 60\%, and 80\%, each task has the same computation and a fixed consumer tail.

Fig.~\ref{fig7}(a) shows that both policies serve all requests at low load. As contention increases, independent release uses idle intervals after compute exit to increase on-time completions. At 40\% load, completions increase from 8/16 to 16/16. Independent release also maintains higher completion capacity at 60\% and 80\% load.

In Fig.~\ref{fig7}(b), reducing the post-compute consumer waiting tail from 32 cycles to zero allows both policies to complete 16/16 requests. This ablation attributes the benefit to slack between computation completion and consumer completion: independent release improves on-time service by eliminating additional occupancy.

\figone{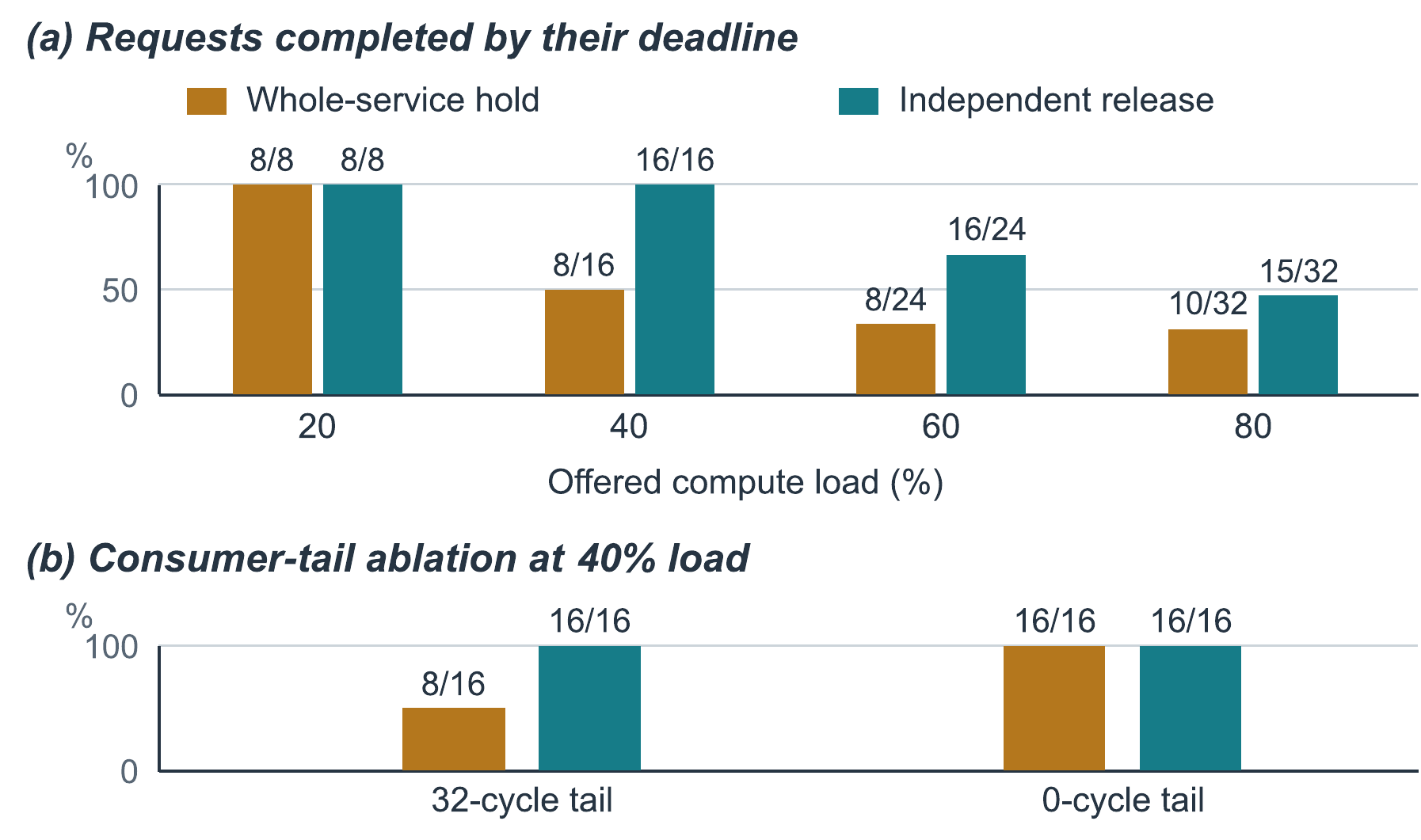}{On-time service under identical arrivals. (a) Whole-service hold and independent release at four offered loads. (b) Reducing the consumer tail from 32 cycles to zero at 40\% load.}{fig7}

\subsubsection{Ablation of Release and Installation Planning}
Fig.~\ref{fig8} compares A0, A1, and A2 on the same Service implementation to distinguish the contributions of resource release and parameter-installation planning.

In the loose scenario, all three complete every request, showing that early release yields an execution benefit only when contention exists. With a consumer tail and a competing request, A0 completes only 1/2 operations, whereas A1 and A2 both complete 2/2. Independent release therefore exploits otherwise idle resources after compute exit.

With additional parameter-update contention, five predefined arrival phases produce 15 requests. A0 and A1 each complete 5/15, while A2 increases completions to 15/15 through joint installation planning. Since A1 and A2 have the same release policy, the difference shows that coordinating installation windows creates additional scheduling opportunities on both sides of a parameter switch.

Fig.~\ref{fig8}(b) explains installation-plan revision. The first old operation's last coefficient read is at 58; the candidate is ready at 24; and the latest installation time is 66. The fixed policy selects the initial earliest installation time, 58, from registered commitments and retains it. A2 retains the legal slots 58--66 and initially selects 58. A second old operation requests admission at 26 and requires old coefficients in cycles 59--62. Although A1 can reuse the read port, installation fixed at 58 is incompatible with these reads, so it safely rejects the request. A2 rechecks and moves installation to 62, protecting the second old operation and performing its last old read and installation on the same edge. The new operation arriving at 70 then reads new coefficients in cycles 101--104 and produces its last post-configuration output at 106. Both policies complete the first old operation and the new-version operation; their completed sets differ only in the second old operation: A1 completes 2/3 and A2 completes 3/3.

For the single-update case, local repair also reaches 3/3. Joint planning adds value when updates compete for a port: with legal slots $\{10,11\}$ and $\{11,12\}$ and an initial plan of 10/11, a new consumer postponing the first update to 11 causes a conflict if its neighbor stays fixed. QCORE instead revises both to 11/12, accepting the consumer while preserving both update deadlines. If both updates are restricted to 11, it rejects the new request and retains existing commitments. Independent coordination-core replay verifies these behaviors. In the model enumeration of 21,984 inputs, QCORE matches the exhaustive reference's feasibility decisions and plans, admitting all 11,496 feasible inputs---328 more than local repair (2.85\% of the feasible set). These are bounded, fixed-order comparisons (Appendix~A).

These experiments show that independent release, installation-window revision, and multiple-update coordination address resource reuse, parameter switching, and shared commit capacity, respectively. Together, they determine which tasks fixed hardware can admit.

\figwide{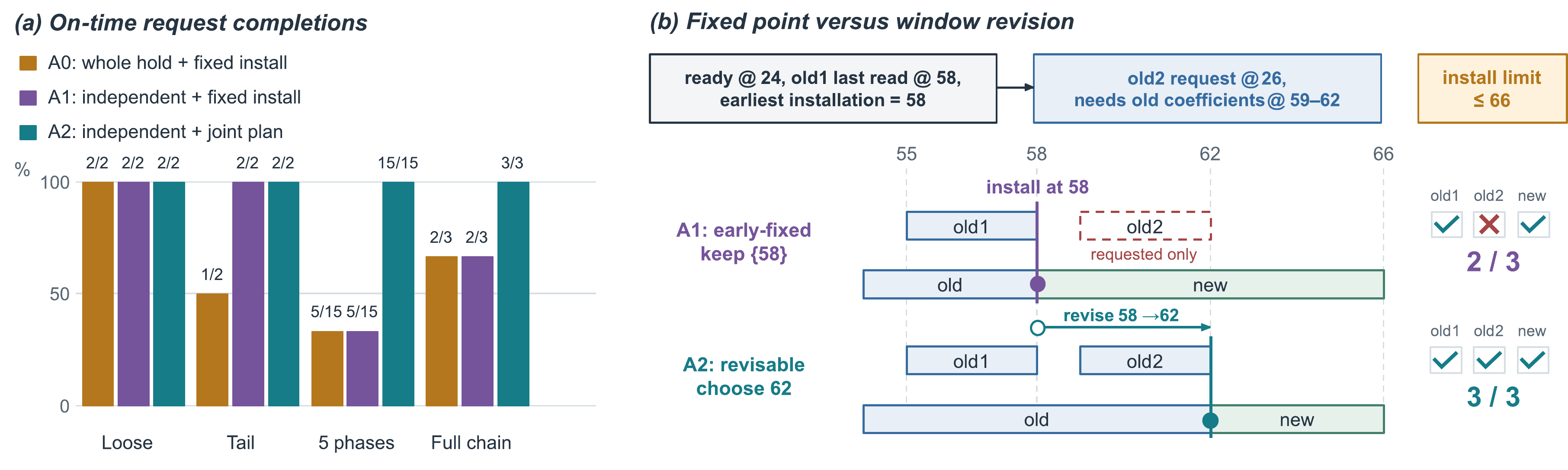}{Same-resource policy ablation. (a) On-time completion fractions for three policies and four request sets; the five-phase set contains 15 requests. (b) Active-coefficient switching and old-consumer reads under common inputs. Operation old1 sets the initial earliest installation time to 58; later-arriving old2 requires old coefficients through 62. A1 retains 58 and rejects old2; A2 moves installation to 62 within deadline 66 and completes all three operations. Dashed lines show the rejected request's read requirements. The last old read on the installation edge still observes the old value; the new value becomes available on the next edge.}{fig8}

\subsection{Cost and Update Tradeoffs of Parameter Protection}
To analyze parameter-lifecycle protection strategies, parameterized RTL endpoints use a common multiply--accumulate datapath. Fig.~\ref{fig9} compares the resource cost and installation latency of value snapshot, version binding, and active configuration across object sizes.

The experiment uses eight in-flight consumers and parameter objects of 32, 128, 512, 2,048, and 8,192 bits. Fig.~\ref{fig9}(a) shows similar costs for small objects. As object size grows, snapshot logic and storage costs rise rapidly because each operation requires its own copy. For an 8,192-bit object, snapshots require 2,285 LUTs, versus approximately 897 and 896 LUTs for version binding and active configuration, respectively.

Fig.~\ref{fig9}(b) compares update-request-to-installation latency. Value snapshot and version binding are mainly limited by candidate loading time. Active configuration must additionally wait for protected old reads to finish. With 8,192 bits and eight consumers, loading takes 512 cycles and waiting takes 4,407 cycles, totaling 4,919 cycles. This metric excludes the additional binding time incurred by per-operation snapshot copying.

Snapshots suit small parameters that can be captured early. As objects or the number of concurrent copies grow, version references reduce copying overhead. Endpoints that directly read active registers preserve consistency through installation windows. The choice depends on object size, the number of in-flight consumers, and access semantics; this experiment characterizes the corresponding cost trends. Finite version capacity also makes reclamation scope consequential. In Appendix~A's two-object model, per-object reclamation completes 2/2 updates with 16 words of parameter images, whereas shared-domain RCU completes 1/2 and needs 20 words to reach 2/2. The capacity difference comes from avoiding retention by unrelated readers; it excludes control metadata.

\figone{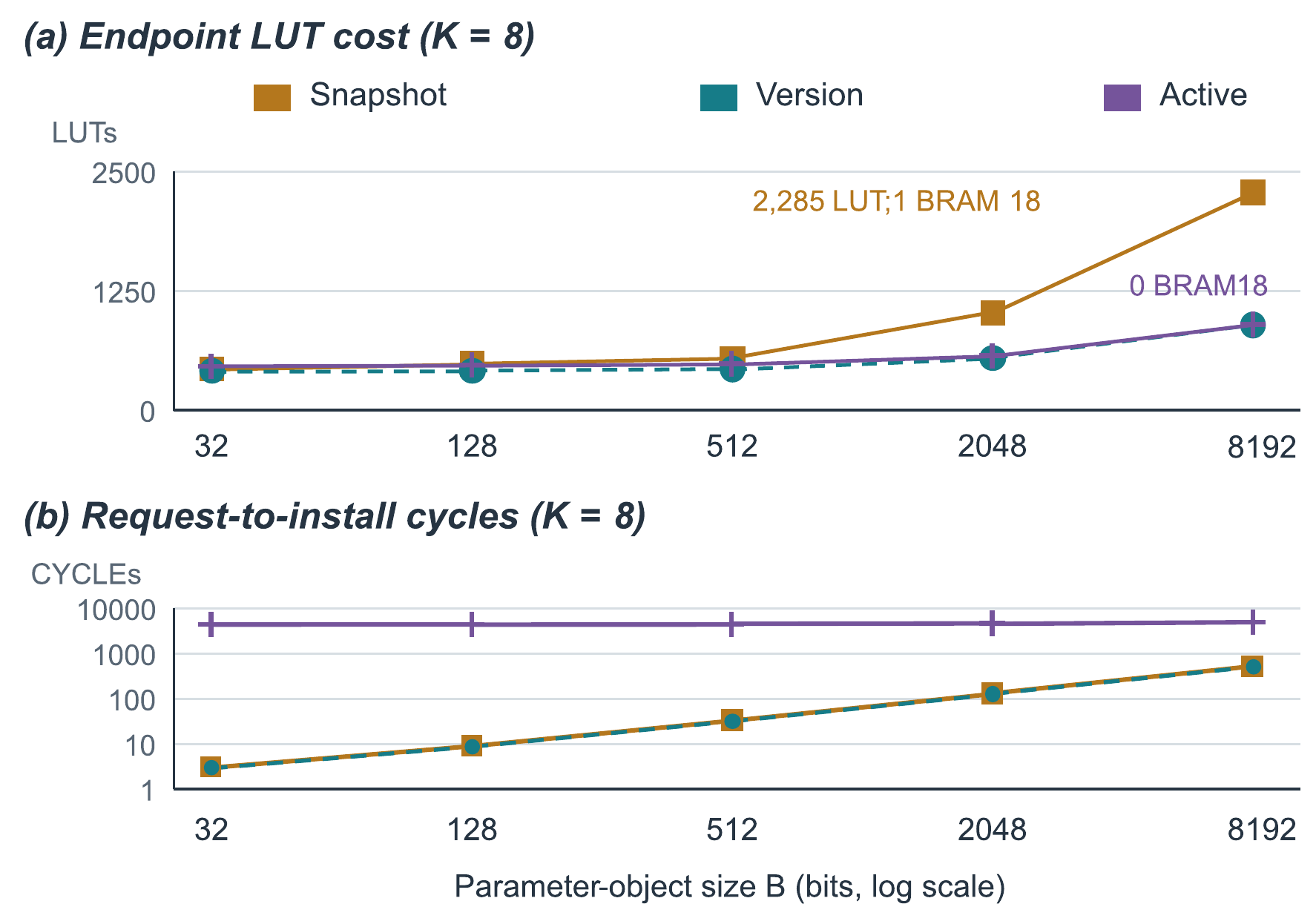}{Cost and installation latency of three parameter-protection modes versus object size at $K=8$. (a) Endpoint LUT usage. At 8,192 bits, value snapshot uses one BRAM18, while version binding and active configuration use none. (b) Update-request-to-installation cycles. The snapshot and version-binding curves overlap; active configuration waits for protected old reads to finish.}{fig9}

\begin{table}[!b]\caption{Digital Service Resources: Policies and Backend Sharing}\label{tab4}\label{tab5}
\centering\footnotesize
\begin{tabular}{lrrrr}\toprule
Configuration & LUT & FF & DSP & B18\\\midrule
A0, $N=24$ & 12,814 & 17,561 & 20 & 22\\
A1, $N=24$ & 12,929 & 17,563 & 20 & 22\\
A2, $N=24$ & 12,951 & 17,565 & 20 & 22\\
$\Delta$ A2$-$A0 & 137 & 4 & 0 & 0\\\midrule
S, $N=24$ & 12,951 & 17,565 & 20 & 22\\
$D_2$, $N=24$ & 14,764 & 18,551 & 33 & 22\\
S, $N=52$ & 16,254 & 27,419 & 20 & 22\\
$D_2$, $N=52$ & 18,859 & 28,520 & 33 & 22\\\bottomrule
\end{tabular}
\par\smallskip\raggedright A0: whole-service hold; A1: independent release; A2: full QCORE. S: one shared NN backend; $D_2$: two static NN backends. $N$: code-block contexts; B18: BRAM18-equivalent usage.
\end{table}

\begin{table*}[!t]\caption{Related Systems: Calibration, QEC, Runtime Coordination, and Scale}\label{tab6}
\centering\footnotesize\renewcommand{\arraystretch}{1.18}\setlength{\tabcolsep}{3pt}
\begin{tabularx}{\textwidth}{@{}P{.15\textwidth}P{.17\textwidth}P{.105\textwidth}P{.115\textwidth}P{.23\textwidth}Y@{}}\toprule
System & Execution & Calibration & QEC & Runtime coordination & Reported scale\\\midrule
QICK \cite{ref8} & PS + tProcessor & Python/PS & Not reported & Timed sequence execution & 1-qubit experiment; 8 DACs/8 ADCs\\\addlinespace[3pt]
QubiC \cite{ref22,ref25,ref26} & Host/ARM + distributed cores & Software & Not reported & Distributed feed-forward & 8-qubit configuration; 3-qubit teleportation\\\addlinespace[3pt]
RISC-Q \cite{ref10} & RISC-V + timed FIFOs; RF modules & On-chip CPU & Extension interface & CPU-driven feedback and updates & 16 DACs/8 ADCs; qubit limit not reported\\\addlinespace[3pt]
RISC-Q QEC \cite{ref24} & Root/leaf nodes & Not reported & Root hardware decoder & Pipelined decoding and correction & Three-board $d=3$ loopback; $d=21$ estimate\\\midrule
\textbf{QCORE} & Management CPU + timed execution & Shared backend & Shared backend & \textbf{Independent release;}\newline\textbf{parameter protection;}\newline\textbf{joint admission} & 52 $d=3$ blocks; 884-qubit digital workload\\\bottomrule
\end{tabularx}
\par\smallskip\raggedright All systems include timed control/readout; QubiC, RISC-Q, RISC-Q QEC, and QCORE also report feedback. PS: processing system. ``Not reported'' refers to the cited work. QCORE shares local backends and supports external compute; Section~V-A defines its validation scope.
\end{table*}

\subsection{Digital Service Implementation Cost and Sharing Benefits}
We report digital Service resources, including admission, identity and parameter management, and endpoints, separating policy increments from shared-versus-static backend savings.

\subsubsection{Digital Service Top Level and Policy Increments}
The complete A2 digital Service in Table~\ref{tab4} uses 12,951 LUTs, 17,565 FFs, 20 DSPs, and 22 BRAM18-equivalent blocks. With safety infrastructure common to all policies, A1 adds 115 LUTs and two FFs over A0, and A2 adds another 22 LUTs and two FFs. The A2--A0 policy increment is 137 LUTs and four FFs, a 1.07\% LUT increase, with unchanged DSP and BRAM usage.

Thus, in the $N=24$ Service build with common parameter protection, identity management, and endpoint checks, changing release and installation policies requires only a small amount of additional control logic.

\subsubsection{Shared Versus Statically Assigned Backends}
Table~\ref{tab5} compares one shared NN backend (S) with two statically assigned backends ($D_2$) under identical workloads, parameter deployment, and external consumption schedules. $D_2$ duplicates NN computation and writable weights for two contexts; S dynamically reuses the backend while meeting the same result-use deadlines.

For both $N=24$ and $N=52$, the shared backend produces the same external results as $D_2$. Relative to the two-backend design, S reduces LUT usage by 12.28\% and 13.81\%, respectively, and reduces DSP usage from 33 to 20, or 39.39\%. BRAM18-equivalent usage remains 22.

At $N=24$, S and $D_2$ finish generating NN results in cycles 13,522 and 13,426, respectively, and both satisfy external consumption at cycle 13,526. The shared backend absorbs internal compute-time differences using delivery slack, converting scheduling flexibility within the deadline into hardware savings.

\section{Discussion}
\subsection{Validated Scale and Scaling Path}
The largest fully validated configuration contains 52 distance-3 rotated surface-code contexts. Each block contains nine data qubits and eight ancilla qubits, corresponding to a digital processing workload for 468 data qubits and 416 ancilla qubits, or 884 physical qubits within the code blocks. This configuration completes 216 input batches, 89,856 NN results, and 3,744 frame-processing operations across 72 three-round windows.

Single-chip scaling requires more in-flight records, parameter and history storage, and sufficient compute throughput and memory bandwidth. Larger code distances also require endpoints that meet the increased data volume, decoding-state, and deadline demands. The lifecycle interface exposes these conditions to admission logic and backend replacement.

Multiple QCORE nodes can distribute local coordination by control domain. The Host deploys tasks and configurations, and each controller handles admission, parameter protection, and reclamation after submission. This resembles the division between Host work submission and device-local execution scheduling in GPUs \cite{ref27}. If each node retains the current processing capacity of 52 code blocks, and workloads can be partitioned while satisfying cross-node dependencies and link-bandwidth requirements, eight nodes correspond to 416 distance-3 blocks, or a digital workload capacity of 7,072 physical qubits. This is a conditional estimate based on the single-node results. General-purpose program request generation, cross-node time synchronization, and global dependency management remain requirements for further integration.

\subsection{Comparison with Related Systems}
Table~\ref{tab6} compares execution, calibration, QEC, coordination, and reported scale. QICK and QubiC provide timed control, readout, and local program execution; QubiC also demonstrates mid-circuit feedback. RISC-Q supports on-chip calibration, with follow-up work integrating distributed QEC. QCORE additionally delegates shared-service admission, parameter protection, installation, and reclamation to hardware, reusing computation before subsequent reads and outputs finish while preserving registered operations and update deadlines.

NVQLink operates at the cooperation layer between controllers and real-time CPUs/GPUs, supporting time-sensitive QEC and calibration \cite{ref6}. QCORE can perform suitable tasks locally or send larger computations to external accelerators, then incorporate returned results according to required versions and deadlines. The benefits of combining them depend on workload and link budgets; the present experiments quantify local sharing and coordination.

Tasks expose occupancy, parameter access, and consumer deadlines. Timed tiles preserve execution timing, shared services exploit result-use slack, and hardware maintains dependencies. Same-resource ablations and Appendix~A distinguish release, reclamation, and plan-revision benefits.

Hardware handles lifecycle events after submission. General-purpose request generation, cross-controller synchronization, and global dependency management remain integration requirements for Host-level program deployment and cross-node organization.

\section{Conclusion}
QCORE coordinates shared classical computation, parameter updates, and deterministic execution through separate resource, parameter, and consumer lifecycles. After submission, hardware checks use conditions, protects parameters through snapshots, versions, or installation windows, and jointly admits reservations and updates. Independent reclamation enables reuse while protecting unfinished operations and registered future consumers.

The 100 MHz implementation reduces Ramsey calibration and QEC replay completion time by 16.83\% through independent release and achieves 15/15 on-time completions under five-phase contention through joint installation planning. Parameter-size experiments characterize copying, version retention, and update waiting. Under identical result deadlines, sharing saves 12.28\%--13.81\% LUTs and 39.39\% DSPs relative to two static NN backends.

The model supports calibration, QEC, and other classical tasks with less fine-grained Host coordination and compute replication. Together with program deployment, synchronization, and external-compute interfaces, it provides a basis for multi-controller Host management and low-latency quantum--classical cooperation.





\clearpage
\appendices
\setcounter{table}{0}
\setcounter{figure}{0}
\renewcommand{\theHfigure}{appendix.\Alph{figure}}
\section{Request Admission Under Limited Resources and Mechanism Comparisons}
\renewcommand{\thesubsection}{\thesection.\arabic{subsection}}
\renewcommand{\thesubsectiondis}{\thesection.\arabic{subsection}.}
\renewcommand{\thetable}{\thesection-\Roman{table}}
\renewcommand{\theHtable}{\thesection.\arabic{table}}
\renewcommand{\thefigure}{\Alph{figure}}
\subsection{Protection Mechanisms and Common Execution Conditions}
This appendix examines how QCORE admits new work under limited resources while preserving existing operations' parameter and deadline requirements. Fig.~\ref{figA} shows two benefits: reclamation based on actual per-object references allows limited version storage to accommodate subsequent updates, and joint revision of installation plans allows a shared port to accept new old-parameter consumers. All comparisons follow the main manuscript's safe-execution contracts and separately examine how version capacity, future reads, and installation contention affect admission.

Comparisons use identical datapaths, identity protection, and timing constraints. With the same capacity, they vary reclamation granularity or the scope of installation-plan revision. Capacity-compensation configurations with extra banks are identified separately. RCU separates version publication from old-object reclamation \cite{ref19,ref28}; shared-domain and object-domain models analyze how reclamation waiting scope affects the outcome. Installation comparisons additionally test whether new requests and existing updates can execute jointly on time. An exhaustive reference uses the same known information and enumerates all installation combinations to independently verify feasibility. Appendix~B analyzes its computational work.

\begin{table}[!ht]\caption{Comparison Mechanisms and Fixed Conditions}\label{tabAI}
\centering\footnotesize
\begin{tabularx}{\columnwidth}{@{}P{.29\columnwidth}YY@{}}\toprule
Comparison & Varied rule & Common conditions\\\midrule
Shared-domain/object-domain reclamation & Scope of reclamation waiting & Binding, last read, asynchronous publication, capacity, deadlines\\
QCORE object protection & Reclamation by actual object references & Same as above\\
Fixed/local/joint installation & Installation times and scope of coordinated changes & Parameter/identity protection, independent release, future reads, order, port\\
Exhaustive enumeration & Enumerate complete installation plans & Same known windows, order, and capacity as joint planning\\\bottomrule
\end{tabularx}\end{table}

\subsection{New-Update Admission Under Finite Version Capacity}
To evaluate how reclamation granularity affects update deadlines and version capacity, the model uses two independent parameter objects, P and Q. Each has two four-word banks and independent read/write ports, transferring one word per cycle. Old readers bind P0 and Q0 at cycle 0. Loading of P1 begins at cycle 3, writes occur in cycles 4--7, and publication occurs at cycle 8. Previously bound P0 readers continue through cycle 16, while the long Q reader continues through cycle 80. P2 is requested at cycle 20, also requires four loading cycles, and must be published by cycle 32. Every scheme uses ideal, immediate detection of reclamation conditions, and actual reads sustain the long references.

Fig.~\ref{figA}(a) shows that QCORE completes both P1 and P2 updates with the same capacity of two P versions. After inactive P0's actual reference count reaches zero at cycle 16, its bank can be loaded again from cycle 17. The P2 loading request is accepted at cycle 20, its four words are written in cycles 21--24, and it is published at cycle 25, before the deadline of 32. Shared-domain RCU waits for existing P and Q readers in the same domain to exit, retaining P0 through cycle 80. Old P0 and active P1 occupy both version slots, preventing P2 from loading on time. P2 is safely rejected at its deadline, so only 1/2 updates complete. Reclamation by actual object references eliminates blocking of P2 by unrelated Q readers, increasing on-time updates to 2/2 at the same capacity.

The lower part of Fig.~\ref{figA}(a) compares version capacities needed for the same 2/2 update-completion capability. Adding a third P bank allows shared-domain RCU to publish P2 at cycle 25 as well. The P/Q parameter images then occupy 20 words, compared with 16 words in QCORE, a 20\% reduction. For P alone, capacity decreases from 12 to eight words. Object-level reclamation makes P0's space available for P2 in time, avoiding simultaneous retention of old P0, active P1, and the P2 image being loaded, and reducing the version capacity required to meet update deadlines.

We further set Q's last read to cycles 16, 32, 80, and 128. QCORE completes the same deadline-constrained updates in all four cases. Shared-domain reclamation incurs no additional retention when Q's last read is at 16; the other three cases incur 16, 64, and 112 cycles of cross-object retention, respectively. All 13 configurations preserve correct old-version reads. Parameter-image capacity counts data words in banks; control and reference metadata and physical memory mapping depend on the implementation.

QCORE also incorporates parameter reads that have not yet occurred into request admission. In Fig.~8(b) of the main manuscript, old2 requests admission at cycle 26 and needs old coefficients in cycles 59--62. Registering these future reads allows installation decisions to retain its required state. Version-addressed endpoints retain references from binding through the actual last read, while active-configuration endpoints incorporate future reads into installation windows. Accepted operations therefore remain protected while waiting in queues, and subsequent updates have explicit switching boundaries.

\subsection{New Consumers and a Shared Installation Port}
To isolate installation scheduling, all four policies share the same version protection, future-consumer registration, independent release, record capacities, and deadlines. Fixed installation retains the original installation times. Local repair exhaustively searches legal slots for the directly affected update while holding other updates fixed. QCORE constructs a joint plan in a fixed order. The independent exhaustive feasibility reference enumerates every combination under the same constraints and chooses the lexicographically earliest feasible solution. The experiment uses one commit port, one cycle per installation, and a common model decision budget.

Fig.~\ref{figA}(b) shows how a new consumer triggers joint revision. Two updates, $U_0$ and $U_1$, have legal slots $\{10,11\}$ and $\{11,12\}$, with an original plan of 10/11. A new old-parameter consumer raises $U_0$'s installation lower bound to 11. Local repair retains $U_1$ at 11, so its 11/11 trial violates the single-port capacity. It rejects the new request and continues the original plan; the conflicting trial is never executed. QCORE moves $U_0$ and $U_1$ jointly to 11/12, admitting the new consumer on the same port while keeping both existing updates on time. The independent exhaustive reference produces the same plan. In the single-update comparison in Fig.~8(b), fixed installation completes 2/3 operations, while local repair, QCORE, and the exhaustive reference all move installation from 58 to 62 and complete 3/3. The multiple-update example further demonstrates the admission benefit of jointly moving neighboring updates.

The benefit comes from jointly exploiting the remaining slack of multiple updates: after $U_0$ moves to 11, $U_1$ can still move to 12 within its deadline. Even when old parameters are protected and every update individually has legal slots, a shared port requires a jointly feasible schedule. Object-domain reclamation specifies when old versions can be released, but that rule alone does not construct the deadline-constrained installation schedule here. Revising only the directly affected update also misses the opportunity to move $U_1$ later. QCORE incorporates new reads, update deadlines, and port capacity into joint admission, expanding the admissible work set while preserving existing commitments.

Tests also cover three boundary cases: the original time remains legal, the neighboring update is already at 12, and new dependencies leave only slot 11 for both updates. In the first two cases, the relevant policies retain or revise the plan. In the last case, all safe policies reject the new consumer and complete the original two installations. Across five scenarios and four policies, all 20 configurations avoid version errors, port conflicts, and violations of existing commitments. The single-update case has a corresponding actual Service RTL implementation. Independent coordination-core replay provides protocol-level evidence for the multiple-update case under the timing conditions in Section~V-A.

\figwide{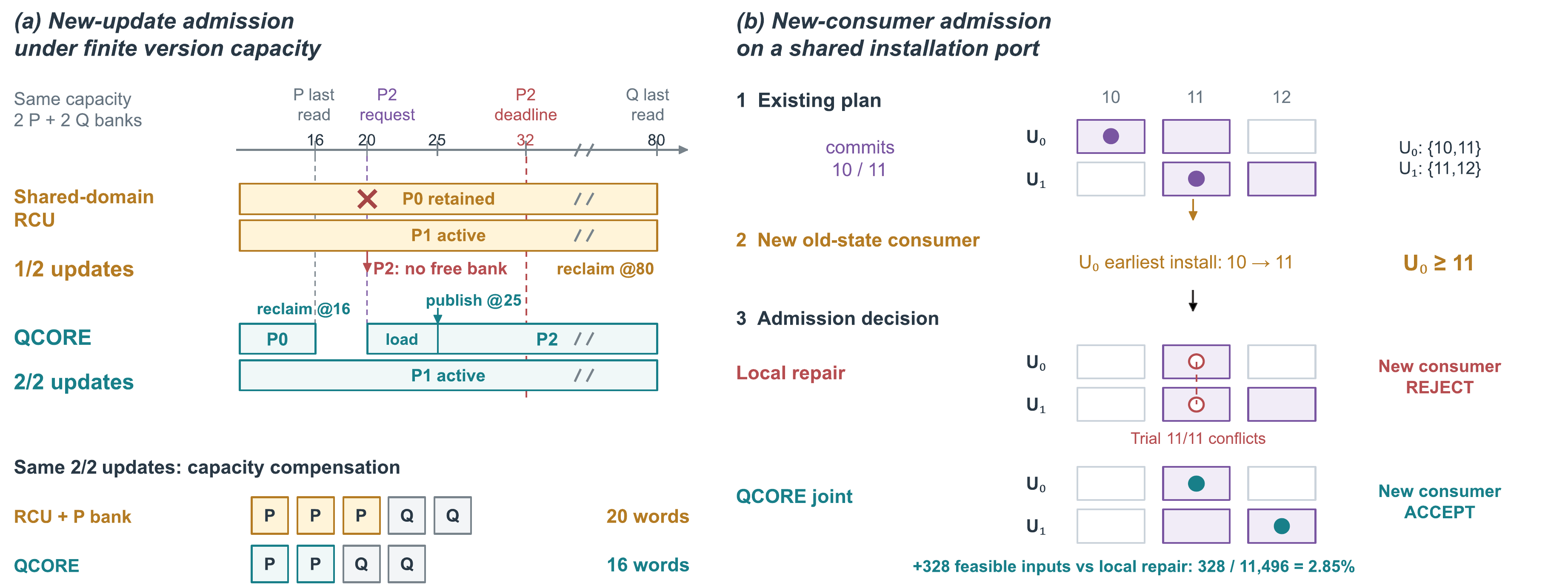}{Admission of new requests under limited resources. (a) With the same version capacity, QCORE reclaims P0 at cycle 16 and publishes P2 at cycle 25, completing 2/2 updates. A long Q reader blocks shared-domain RCU, which completes 1/2. One additional P bank allows shared-domain RCU to complete 2/2, but requires 20 words of parameter images versus QCORE's 16. (b) After a new consumer tightens $U_0$'s window, local repair rejects the 11/11 trial and the new request, retaining the original 10/11 plan. QCORE jointly replans to 11/12 and admits the request. Both policies keep existing updates on time. Both panels use executable models with identical constraints.}{figA}

\subsection{Finite Enumeration and Feasibility Coverage}
To determine whether these differences extend beyond an individual example, we enumerate all nonempty initial legal sets for two to three updates over three to four slots, including sets with holes. For inputs with a feasible old plan, we use the earliest feasible old plan, then select each update in turn and tighten its lower bound while leaving all other conditions unchanged. This yields 21,984 inputs: 11,496 remain feasible after tightening and 10,488 are infeasible. The local policy is also checked against an independent enumeration that holds all other installation times fixed.

\begin{table}[!t]\caption{Completion and Feasibility Results of Installation Policies}\label{tabAII}
\centering\footnotesize
\begin{tabularx}{\columnwidth}{@{}P{.28\columnwidth}YYY@{}}\toprule
Policy & One update: operations completed & Two updates: new consumer admitted & Enumeration: feasible inputs admitted\\\midrule
Fixed installation & 2/3 & 0/1 & 10,488\\
Local repair & 3/3 & 0/1 & 11,168\\
QCORE joint planning & 3/3 & 1/1 & 11,496\\
Exhaustive reference & 3/3 & 1/1 & 11,496\\\bottomrule
\end{tabularx}\end{table}

Among the 11,496 inputs that remain feasible, QCORE admits all of them: 328 more than local repair (2.85\%) and 1,008 more than fixed installation. Joint planning and the exhaustive reference agree on both feasibility and the selected plan. Under a fixed order, one port, and one-cycle installations, joint revision fully uses other updates' remaining slots and avoids conservative rejection caused by fixed neighboring installations. The 2.85\% is the admission difference within this enumerated set; application benefits depend on request distributions, port contention, and consumer slack.

These comparisons demonstrate QCORE's admission benefits in terms of version capacity and a shared installation port. Reclamation by actual object references reduces occupancy caused by unrelated readers. Future-consumer registration protects queued operations' parameter requirements. Joint installation planning uses slack across updates so that new requests and existing commitments can execute together. QCORE connects these conditions through unified admission and actual endpoint events, preserving parameter consistency and execution deadlines while enabling early resource reuse.

\section{Feasibility and Computational Work of Joint Installation Planning}
\setcounter{table}{0}
\subsection{Feasibility Under a Fixed Order}
Equations~(1)--(3) in the main manuscript define resource constraints, consumer deadlines, and installation windows. We now prove the feasibility of sequential earliest selection under a fixed update order. Each installation occupies one cycle on a single port, with legal sets $W_1,\ldots,W_m$. Choose $T_1=\min W_1$, then
\begin{equation*}
T_i=\min\{t\in W_i\mid t\ge T_{i-1}+1\}.
\end{equation*}
If the set is empty at any step, no feasible schedule exists for that fixed order.

For any feasible sequence $F$ with the same order, initially $T_1\le F_1$. If $T_{i-1}\le F_{i-1}$, then $F_i\ge F_{i-1}+1\ge T_{i-1}+1$, and $F_i\in W_i$. The earliest choice therefore exists at this step and satisfies $T_i\le F_i$. By induction, the method constructs a schedule whenever a feasible solution exists in that order. This conclusion assumes a fixed order, one-cycle installations, and determined legal sets. Different orders, multiple ports, or variable installation lengths require appropriate scheduling models.

The current window representation is a 64-bit bitmap of discrete slots, and legal sets may contain holes. After a proposal is formed, actual exit, retirement, and installation events immediately update the active ledger and its revision number. Stale proposals must be rechecked, and final registration verifies time and revision again. Mathematical feasibility is therefore conditional on the validity of the state snapshot used; the hardware protocol maintains that condition through actual authorization.

To test this construction and its rejection conditions, Appendix~A enumerates 21,984 inputs with two to three updates and three to four slots, including legal sets with holes and lower-bound tightening for one update. Joint planning and the independent exhaustive reference agree on feasibility and the chosen plan. Additional checks cover 4,096 three-window combinations over four slots and 61,440 tightening cases, with 33,792 accepted and 27,648 rejected. Together with the proof, these finite-model checks cover the stated fixed-order conditions. RTL regression supplies corresponding evidence for actual endpoints and event handling.

\subsection{Computational Work Compared with Exhaustive Enumeration}
Agreement on feasibility shows that joint planning finds the same schedule as the reference; computational work determines whether such checking can be used frequently at runtime. Suppose the $m$ updates have $k_1,\ldots,k_m$ legal slots. The exhaustive reference traverses the Cartesian product of these sets, yielding $\prod_i k_i$ complete candidate vectors, or $k^m$ if every update has $k$ candidates. It checks each vector for order and port constraints. Sequential earliest selection scans ordered legal sets one by one and checks at most $\sum_i k_i$ candidate slots per construction. For bitmaps of length $H$, direct serial scanning requires at most $mH$ bit checks. These counts assume that legal sets have already been formed and update order is fixed.

Table~\ref{tabBI} gives feasible examples where every candidate set is $\{0,\ldots,7\}$. Sequential earliest selection visits $1,2,\ldots,m$ candidates and constructs the plan $0,1,\ldots,m-1$. The exhaustive reference still traverses all combinations to independently confirm the earliest solution. As update count increases from two to eight, exhaustive candidate vectors increase from 64 to 16,777,216, while sequential candidate checks increase from three to 36. Combination counts are calculated analytically, and sequential-check counts are verified using the same reference model. The table characterizes algorithmic work.

\begin{table}[!t]\caption{Plan-Construction Work with Identical Legal Sets}\label{tabBI}
\centering\footnotesize
\begin{tabularx}{\columnwidth}{@{}YYYY@{}}\toprule
Updates $m$ & Candidates per update & Exhaustive: complete vectors & Sequential: slot checks\\\midrule
2 & 8 & 64 & 3\\
4 & 8 & 4,096 & 10\\
8 & 8 & 16,777,216 & 36\\\bottomrule
\end{tabularx}\end{table}

One exhaustive check examines an entire installation vector; one sequential-selection check examines a candidate slot. The current model assigns the same decision budget to every policy to isolate plan-selection capability. Hardware control latency also includes window formation, state validation, and registration. QCORE exploits the structure of the fixed-order problem to avoid combinatorial enumeration, preserving feasibility coverage under identical constraints and providing a basis for a bounded hardware implementation.

\section{Hardware State, Events, and Interface Contracts}
\setcounter{table}{0}
\subsection{Hardware Control After Request Submission}
Service Admission (the request adapter between Admission FSM and service endpoints) uses five states: IDLE, SEND, WAIT, APPLY, and ACK. They handle idle reception, submission to the ledger, waiting for authorization, post-authorization checking and binding, and endpoint acknowledgment, respectively. Effect Admission (the decision core within Admission FSM) uses IDLE, SELECT, CHOOSE, ENGINE, UPDATE, RETRY, and PUBLISH to form proposals for resources and installation plans. Commitment Ledger (the reservation and dependency records and registration control within Admission FSM) uses IDLE, CERT, WAIT, MAINT, and REGISTER for plan certification, proposal maintenance, and final registration. Core proposal, ledger authorization, and endpoint acceptance are distinct boundaries. Fig.~5(a) combines the decision core and ledger into an overall flow; PROPOSE summarizes proposal formation and publication. The separate states listed here describe the actual implementation.

Exit, retirement, and actual installation immediately update the active ledger and its revision number. A proposal carries the revision used during checking. Final publication and registration both validate freshness; invalidated proposals are rechecked, and tentative plans do not modify existing commitments. After ledger registration, the adapter proceeds with parameter binding. If cancellation, a version change, or endpoint rejection prevents actual acceptance, it sends exit and retirement cleanup events tagged with the complete identity.

\subsection{Parameter State and Release Events}
Versioned banks are managed by flags and fields including \texttt{active}, \texttt{valid}, \texttt{loading}, \texttt{word mask}, \texttt{epoch}, and \texttt{refs}, without a separate enumerated FSM. Overwriting can begin only in inactive, unreferenced banks, and publication requires a complete load. Binding on the same edge as publication obtains the pre-edge version. Active configuration retains active/shadow coefficients, a loading mask, update identity, and deadline. Once prepared, it waits for the ledger plan and installs only when the scheduled time arrives and old references are safe. Endpoint readiness and external plan state jointly define the control flow. In Fig.~4(b), IDLE, LOAD, and PREP. summarize idle, loading, and candidate-ready phases, respectively.

\begin{table}[!t]\caption{Actual Events and Their Effects}\label{tabCI}
\centering\footnotesize
\begin{tabularx}{\columnwidth}{@{}P{.26\columnwidth}YY@{}}\toprule
Event & Record released & Subsequent activity\\\midrule
Resource exit & Resource reservation & Parameter reads, result pipeline, identity responses\\
Object last read & Reference to that object & Other object reads, subsequent output\\
Installation acknowledgment & Corresponding update record & Subsequent pipelines of old/new operations\\
Last output and retirement response & Consumer/identity records drain progressively & Other concurrent operations\\\bottomrule
\end{tabularx}\end{table}

The main coefficient datapath reads \texttt{active\_gain} and \texttt{active\_bias} at consume edge $t$, computes the product, and registers the bias. It forms the multiply--accumulate result at $t+1$ and outputs at $t+2$. The last waveform-bank read, last active-coefficient read, and last output occur separately and cannot be merged into one release signal. Matching the operation ID, generation, and relevant resource identity prevents late events from cleaning up a new owner.

\subsection{Error Reporting and Result Acceptance}
Active Config (the endpoint containing Configuration registers and associated coefficient-processing and version-checking logic) compares \texttt{expected\_epoch} with the actual epoch at the read edge. On a mismatch, it sets a sticky \texttt{fault} and outputs \texttt{out\_error} aligned with data, identity, actual version, and first/last flags. The \texttt{out\_valid} signal continues to indicate that an output sample is present. The current Service boundary exports these signals; application consumers should use data only when valid and error-free. Atomic acceptance of an entire operation additionally requires accumulated operation-error state and buffering or deferred result commitment.

Last-read and last-output events must still release references and drain identities; cleanup does not depend on whether results were accepted for use. The current connections have been checked through static path inspection, and normal endpoint outputs through regression. Error-flag propagation and application-level result acceptance are separate interface responsibilities. Results returned from external computation likewise require identity, deadline, and parameter contracts.


\begin{thebibliography}{35}
\bibitem{guoVLSI26} Y. Guo \emph{et al.}, ``A cryo-CMOS fully-integrated superconducting qubit reflectometry readout IC with FCNN-based adaptive drift mitigation and FIR-DSM-based stimulus TX,'' in \emph{Proc. IEEE/JSAP Symp. VLSI Technol. Circuits}, 2026, pp. 1--3, doi: \nolinkurl{10.1109/VLSITechnologyandCir65830.2026.11577211}.
\bibitem{liASIC26} H. Li \emph{et al.}, ``A digital baseband ASIC targeting 99.999\% fidelity for qubit control and readout with ROI-based classification,'' in \emph{Proc. IEEE Int. Symp. Circuits Syst. (ISCAS)}, 2026, pp. 3126--3130, doi: \nolinkurl{10.1109/ISCAS66217.2026.11562862}.
\bibitem{liCal26} H. Li, Q. Liu, Y. Guo, T. Li, Z. Wang, and H. Jiang, ``A fewer-cycles, resource-efficient concurrent calibration method for compact transmon qubit control and readout circuit,'' in \emph{Proc. IEEE Int. Symp. Circuits Syst. (ISCAS)}, 2026, pp. 3116--3120, doi: \nolinkurl{10.1109/ISCAS66217.2026.11562708}.
\bibitem{ref1} A. Cross \emph{et al.}, ``OpenQASM 3: A broader and deeper quantum assembly language,'' \emph{ACM Trans. Quantum Comput.}, vol. 3, no. 3, Sep. 2022, Art. no. 12, doi: \nolinkurl{10.1145/3505636}.
\bibitem{ref2} X. Fang \emph{et al.}, ``CaliQEC: In-situ qubit calibration for surface code quantum error correction,'' in \emph{Proc. 52nd Annu. Int. Symp. Comput. Archit. (ISCA)}, 2025, pp. 1402--1416, doi: \nolinkurl{10.1145/3695053.3731042}.
\bibitem{ref3} Google Quantum AI and Collaborators, ``Quantum error correction below the surface code threshold,'' \emph{Nature}, vol. 638, no. 8052, pp. 920--926, Feb. 2025, doi: \nolinkurl{10.1038/s41586-024-08449-y}.
\bibitem{ref4} X. Fu \emph{et al.}, ``An experimental microarchitecture for a superconducting quantum processor,'' in \emph{Proc. 50th Annu. IEEE/ACM Int. Symp. Microarchit. (MICRO)}, 2017, pp. 813--825, doi: \nolinkurl{10.1145/3123939.3123952}.
\bibitem{ref5} X. Fu \emph{et al.}, ``eQASM: An executable quantum instruction set architecture,'' in \emph{Proc. IEEE Int. Symp. High-Perform. Comput. Archit. (HPCA)}, 2019, pp. 224--237, doi: \nolinkurl{10.1109/HPCA.2019.00040}.
\bibitem{guoISSCC23} Y. Guo \emph{et al.}, ``A polar-modulation-based cryogenic qubit state controller in 28nm bulk CMOS,'' in \emph{Proc. IEEE Int. Solid-State Circuits Conf. (ISSCC)}, 2023, pp. 508--510, doi: \nolinkurl{10.1109/ISSCC42615.2023.10067495}.
\bibitem{guoISSCC24} Y. Guo \emph{et al.}, ``29.4 A cryo-CMOS quantum computing unit interface chipset in 28nm bulk CMOS with phase-detection based readout and phase-shifter based pulse generation,'' in \emph{Proc. IEEE Int. Solid-State Circuits Conf. (ISSCC)}, 2024, vol. 67, pp. 476--478, doi: \nolinkurl{10.1109/ISSCC49657.2024.10454392}.
\bibitem{ref6} S. A. Caldwell \emph{et al.}, ``Platform architecture for tight coupling of high-performance computing with quantum processors,'' 2025, arXiv:2510.25213. [Online]. Available: \url{https://arxiv.org/abs/2510.25213}
\bibitem{ref7} M. Zhang \emph{et al.}, ``Exploiting different levels of parallelism in the quantum control microarchitecture for superconducting qubits,'' in \emph{Proc. 54th Annu. IEEE/ACM Int. Symp. Microarchit. (MICRO)}, 2021, pp. 898--911, doi: \nolinkurl{10.1145/3466752.3480116}.
\bibitem{ref8} L. Stefanazzi \emph{et al.}, ``The QICK (Quantum Instrumentation Control Kit): Readout and control for qubits and detectors,'' \emph{Rev. Sci. Instrum.}, vol. 93, no. 4, Apr. 2022, Art. no. 044709, doi: \nolinkurl{10.1063/5.0076249}.
\bibitem{ref9} Y. Xu \emph{et al.}, ``QubiC: An open-source FPGA-based control and measurement system for superconducting quantum information processors,'' \emph{IEEE Trans. Quantum Eng.}, vol. 2, pp. 1--11, 2021, doi: \nolinkurl{10.1109/TQE.2021.3116540}.
\bibitem{ref10} J. Liu, Y. Lee, H. Deng, C. Clayton, G. Yang, and X. Wu, ``RISC-Q: A generator for real-time quantum control system-on-chips compatible with RISC-V,'' 2025, arXiv:2505.14902. [Online]. Available: \url{https://arxiv.org/abs/2505.14902}
\bibitem{skoric23} L. Skoric, D. E. Browne, K. M. Barnes, N. I. Gillespie, and E. T. Campbell, ``Parallel window decoding enables scalable fault tolerant quantum computation,'' \emph{Nat. Commun.}, vol. 14, no. 1, Nov. 2023, Art. no. 7040, doi: \nolinkurl{10.1038/s41467-023-42482-1}.
\bibitem{ref11} A. G. Fowler, M. Mariantoni, J. M. Martinis, and A. N. Cleland, ``Surface codes: Towards practical large-scale quantum computation,'' \emph{Phys. Rev. A}, vol. 86, no. 3, Sep. 2012, Art. no. 032324, doi: \nolinkurl{10.1103/PhysRevA.86.032324}.
\bibitem{ref12} L. Caune \emph{et al.}, ``Demonstrating real-time and low-latency quantum error correction with superconducting qubits,'' \emph{Nat. Commun.}, vol. 17, no. 1, Jun. 2026, Art. no. 7383, doi: \nolinkurl{10.1038/s41467-026-73331-6}.
\bibitem{afs22} P. Das \emph{et al.}, ``AFS: Accurate, fast, and scalable error-decoding for fault-tolerant quantum computers,'' in \emph{Proc. IEEE Int. Symp. High-Perform. Comput. Archit. (HPCA)}, 2022, pp. 259--273, doi: \nolinkurl{10.1109/HPCA53966.2022.00027}.
\bibitem{kelly16} J. Kelly \emph{et al.}, ``Scalable in situ qubit calibration during repetitive error detection,'' \emph{Phys. Rev. A}, vol. 94, no. 3, Sep. 2016, Art. no. 032321, doi: \nolinkurl{10.1103/PhysRevA.94.032321}.
\bibitem{ryan17} C. A. Ryan, B. R. Johnson, D. Rist{\`e}, B. Donovan, and T. A. Ohki, ``Hardware for dynamic quantum computing,'' \emph{Rev. Sci. Instrum.}, vol. 88, no. 10, Oct. 2017, Art. no. 104703, doi: \nolinkurl{10.1063/1.5006525}.
\bibitem{ref22} N. Fruitwala \emph{et al.}, ``Distributed architecture for FPGA-based superconducting qubit control,'' 2024, arXiv:2404.15260. [Online]. Available: \url{https://arxiv.org/abs/2404.15260}
\bibitem{ref26} Y. Xu \emph{et al.}, ``QubiC 2.0: An extensible open-source qubit control system capable of mid-circuit measurement and feed-forward,'' 2023, arXiv:2309.10333. [Online]. Available: \url{https://arxiv.org/abs/2309.10333}
\bibitem{ref13} G. Huang \emph{et al.}, ``Updated QubiC: Improved scalability, performance, and QPU support,'' in \emph{Proc. IEEE Int. Conf. Quantum Comput. Eng. (QCE)}, 2024, vol. 2, pp. 612--613, doi: \nolinkurl{10.1109/QCE60285.2024.10430}.
\bibitem{ref14} Y. Zhao \emph{et al.}, ``Distributed-HISQ: A distributed quantum control architecture,'' in \emph{Proc. 58th IEEE/ACM Int. Symp. Microarchit. (MICRO)}, 2025, pp. 564--578, doi: \nolinkurl{10.1145/3725843.3756048}.
\bibitem{ref15} M-Labs. ``ARTIQ real-time I/O concepts.'' \emph{ARTIQ Manual}. Accessed: Oct. 27, 2026. [Online]. Available: \url{https://m-labs.hk/artiq/manual/rtio.html}
\bibitem{ref16} S. Maurya, A. Molavi, A. Albarghouthi, and S. Tannu, ``A case for elastic quantum error correction decoders,'' in \emph{Proc. 21st Eur. Conf. Comput. Syst. (EuroSys)}, 2026, pp. 514--531, doi: \nolinkurl{10.1145/3767295.3803584}.
\bibitem{ref17} X. Luo, J. Song, and X. Su, ``Time-deterministic and instruction-dynamic execution: A field-programmable gate array quantum-control processor,'' \emph{Rev. Sci. Instrum.}, vol. 97, no. 10, Oct. 2026, Art. no. 103204, doi: \nolinkurl{10.1063/5.0356629}.
\bibitem{ref18} J. E. Smith and A. R. Pleszkun, ``Implementing precise interrupts in pipelined processors,'' \emph{IEEE Trans. Comput.}, vol. 37, no. 5, pp. 562--573, May 1988, doi: \nolinkurl{10.1109/12.4607}.
\bibitem{ref19} M. Desnoyers, P. E. McKenney, A. S. Stern, M. R. Dagenais, and J. Walpole, ``User-level implementations of read-copy update,'' \emph{IEEE Trans. Parallel Distrib. Syst.}, vol. 23, no. 2, pp. 375--382, Feb. 2012, doi: \nolinkurl{10.1109/TPDS.2011.159}.
\bibitem{ref28} ``What is RCU?---`Read, Copy, Update'.'' \emph{The Linux Kernel Documentation}. Accessed: Oct. 27, 2026. [Online]. Available: \url{https://docs.kernel.org/RCU/whatisRCU.html}
\bibitem{ref20} C. L. Liu and J. W. Layland, ``Scheduling algorithms for multiprogramming in a hard-real-time environment,'' \emph{J. ACM}, vol. 20, no. 1, pp. 46--61, Jan. 1973, doi: \nolinkurl{10.1145/321738.321743}.
\bibitem{ref27} NVIDIA. ``Programming model.'' \emph{CUDA Programming Guide}. Accessed: Oct. 27, 2026. [Online]. Available: \url{https://docs.nvidia.com/cuda/cuda-programming-guide/01-introduction/programming-model.html}
\bibitem{ref25} Y. Xu \emph{et al.}, ``Automatic qubit characterization and gate optimization with QubiC,'' \emph{ACM Trans. Quantum Comput.}, vol. 4, no. 1, pp. 1--12, Mar. 2023, doi: \nolinkurl{10.1145/3529397}.
\bibitem{ref24} J. Liu, Y. Lee, Y. Xu, G. Huang, and X. Wu, ``A scalable open-source QEC system with sub-microsecond decoding-feedback latency,'' 2026, arXiv:2603.16203. [Online]. Available: \url{https://arxiv.org/abs/2603.16203}
\end{thebibliography}
\end{document}